\documentclass[twocolumn,showpacs,preprintnumbers,amsmath,amssymb]{revtex4}

\usepackage{graphicx}
\usepackage{dcolumn}
\usepackage{bm}
\usepackage{subfigure}

\begin{document}

\title{Thermocapillary migration and interactions of two nondeformable droplets}

\author{Z. Yin}
\email{zhaohua.yin@imech.ac.cn}

\author{L. Chang}
\author{W. Hu}
\author{P. Gao}

\affiliation{National Microgravity Laboratory, Institute of Mechanics, Chinese Academy of Sciences, Beijing 100190,
P.R.China}

\date{\today}

\begin{abstract}

A numerical study on interactions of two spherical drops in
thermocapillary migration in microgravity is presented. Finite-difference methods were adopted and
the interfaces of drops were captured by the front-tracking technique. It is found that the
arrangement of drops directly influences their migrations and interaction, and that the motion of
one drop is mainly determined by the disturbed temperature field because of the existence of the
other drop.
\end{abstract}


\maketitle

\section*{Introduction}
Under the microgravity condition, the thermocapillary migration of droplets or bubbles in matrix liquid is caused by
 the nonuniform interface tension introduced by the temperature gradient. This motion is of great importance in material
  processing and other applications in space. The original work in this field was performed by Young \emph{et al.}\cite{Young:}.
   In their study, the inertial convection and thermal convection are neglected (the so-called $YGB$ Model), and the derived migration velocity is
\begin{equation}
 V_{YGB}=\frac{2U}{(2+3\mu_d/\mu_b)(2+k_{d}/k_{b})} \label{eq1}.
\end{equation}
Here, $U$ is the reference velocity defined by the balance of thermocapillary force and viscosity force on the drop/bubble:
\[U=|\sigma_T||\nabla T_\infty|a/\mu_b,\]
$\mu$ is the kinematic viscosity, $k$ the thermal conductivity, $\sigma_T$ the rate of change of
interfacial tension with temperature, $\nabla T_\infty$ the temperature gradient imposed on matrix
liquid, and $a$ the radius of the drop or bubble. The symbols with the subscript $d$ mean the
parameters of the droplet/bubble, and those of the bulk liquid are indicated by the subscript $b$.

After YGB, there are many other studies on the thermocapillary motion of isolated drop/bubble (see
\cite{Yin:} and references therein). In practice, it is common to have two or more
drops/bubbles in the continues phase, so it is necessary to study their interactions. The first
axisymmetric investigation of two thermocapillary bubbles was conducted by Meyyappan \emph{et
al.}\cite{Meyyappan:}, using the bipolar coordinate. It was found that the smaller bubble
always moves faster than the isolated drop while the bigger one moves slightly slower. Meyyappan
and Subramanian\cite{Meyyappan2:} extended the above work to arbitrarily placed bubbles.
Balasubramaniam and Subramanian\cite{Balasubramaniam:} assumed that two bubbles migrated in
the potential flow (namely, the related Re number is very large), and the matched asymptotic
analysis was adopted to solve the energy equation with large Ma numbers. It was found that the
thermal wake of the leading bubble will disturb the temperature field around the trailing bubble
and reduce its velocity. Interactions between two spherical droplets were firstly studied by
Anderson with a reflection method\cite{Anderson:}. It was found that interactions between
droplets driven by thermocapillary effects are much weaker than those of sedimentation. Ken and
Chen\cite{Ken:} analyzed the axisymmetric motion of two droplets in the bishperical
coordinate, and their later combined analytical-numerical study was about a finite chain of
spherical droplets along the line of their centers \cite{Ken2:}. Interactions of two
deformable droplets in the axisymmetric coordinate were studied by Zhou and Davis\cite{Zhou:}.
Thermocapillary interactions of droplets or bubbles toward a hot wall at finite Reynolds and
Marangoni numbers were numerically studied by Nas \emph{et al.}\cite{Nas:,Nas2:}. It was found
that bubbles and light drops line up perpendicular to the temperature gradient and are evenly
spaced in the horizontal direction. A space experiment observed that a small leading drop could
retard the movement of the big trailing drop \cite{Balasubramaniam2:}.

So far as we know, there is no systematic study on the interaction of two arbitrarily placed drops
in the thermocapillary research, and it will be the main subject here. We focus our investigation
on two droplets with equal size and the same physical parameters (kinematic viscosity, thermal
diffusivity, density, and specific heat). The governing equations and numerical methods will be
introduced in the next section, detailed numerical models and parameters of simulations are in
section $2$, and the results when the inertia and thermal convection are ignorable or not, are
discussed in section $3$ and $4$, respectively.

\section{Governing equations and numerical methods}
In the thermocapillary motion, the two droplets with the same radius $a$ are surrounded by the bulk
fluid in a rectangular box $\Omega=[x_0, x_1]\times[y_0, y_1]\times[z_0, z_1]$ (Fig.
\ref{fig:Model}). The box is closed by no-slip walls. The direction of the temperature gradient is
along the $z$ axis, and $x=0$ is treated in the drop centers. The governing equations for this
problem are:
\begin{eqnarray*}
&& \nabla\cdot \textbf{u}=0, \label{eq2}\\
&&  \frac{\partial {(\rho \textbf{u})}}{\partial
{t}}+\nabla\cdot(\rho \textbf{u} \textbf{u})=-\nabla
p+\nabla\cdot(\mu(\nabla \textbf{u}+\nabla^T
  \textbf{u}))+\emph{\textbf{F}}_{\sigma}, \label{eq3}\\
  && \rho C_p(\frac{\partial T}{\partial {t}}+\textbf{u}\cdot\nabla
  T)=\nabla\cdot(\emph{k}\nabla T). \label{eq4}
\end{eqnarray*}
Here, $\textbf{u}=(u, v, w)$, $\textbf{x}=(x, y, z)\in \Omega$. $\emph{\textbf{F}}_{\sigma}$ is the
body force term calculated by integrating the surface tension across the interface\cite{Yin:}.
Except the different material parameters for the drop phase and the bulk phase, the conservative
equations above are valid for both phases. We define the nondimensional quantities as:
\begin{eqnarray}
&& \bar{\textbf{u}}=\textbf{u}/U,\quad \bar{\textbf{x}}=\textbf{x}/a, \quad \bar{t}=t/(\frac{a}{U}), \nonumber \\
&& \bar{p}=p/(\rho_b U^2), \quad  \bar{T}=T/(|\nabla T_{\infty}|a),\quad \bar{\rho}=\rho/\rho_b, \label{eq5} \\
&& \bar{\mu}=\mu/\mu_b, \quad \bar{k}=k/k_{b}, \quad \bar{C_p}=C_p/C_{pb}, \nonumber \\
&& \bar{\textbf{F}}_{\sigma}={\emph{\textbf{F}}_{\sigma}}a/(\rho_1U^2),\quad
Re=Ua/\nu_b,\quad Ma=Ua/\kappa_b. \nonumber
\end{eqnarray}
Here, $\nu_b=\mu_b/\rho_b$ is the kinematic viscosity, and $\kappa_b = {k_b}/(\rho_b C_{pb})$ the thermal diffusivity of the matrix liquid.
The nondimensional equations can be written as:
\begin{eqnarray}
&& \nabla\cdot \bar{\textbf{u}}=0, \label{eq6}\\
&&  \frac{\partial {(\bar{\rho} \bar{\textbf{u}})}}{\partial
{\bar{t}}}+\nabla\cdot(\bar{\rho} \bar{\textbf{u}}
\bar{\textbf{u}}) \nonumber \\
&& \quad  =-\nabla
\bar{p}+\frac{1}{Re}\nabla\cdot(\bar{\mu}(\nabla
\bar{\textbf{u}}+\nabla^T
  \bar{\textbf{u}}))+\bar{\emph{\textbf{F}}_{\sigma}}, \label{eq7}\\
  && \bar{\rho} \bar{C_p}(\frac{\partial \bar{T}}{\partial {t}}+\bar{\textbf{u}}\cdot\nabla
  \bar{T})=\frac{1}{Ma}\nabla\cdot(\bar{\emph{k}}\nabla \bar{T}). \label{eq8}
\end{eqnarray}
The boundary conditions for velocities are:
\begin{eqnarray}
\begin{aligned}
&\bar{u}|_{\bar{x}=\bar{x}_0,\bar{x}_1}=\bar{v}|_{\bar{x}=\bar{x}_0,\bar{x}_1}=\bar{w}|_{\bar{x}=\bar{x}_0,\bar{x}_1}=0,
 \\
&\bar{u}|_{\bar{y}=\bar{y}_0,\bar{y}_1}=\bar{v}|_{\bar{y}=\bar{y}_0,\bar{y}_1}=\bar{w}|_{\bar{y}=\bar{y}_0,\bar{y}_1}=0,
\\
&\bar{u}|_{\bar{z}=\bar{z}_0,\bar{z}_1}=\bar{v}|_{\bar{z}=\bar{z}_0,\bar{z}_1}=\bar{w}|_{\bar{z}=\bar{z}_0,\bar{z}_1}=0.
 \end{aligned}
\end{eqnarray}
For energy equation, the \emph{Dirichlet} boundary condition is adopted:
\begin{eqnarray}
\begin{aligned}
&\bar{T}|_{\bar{x}=\bar{x_0}}=\bar{T_0}+\bar{z},\quad\bar{T}|_{\bar{x}=\bar{x_1}}=\bar{T_0}+\bar{z},
\\&\bar{T}|_{\bar{y}=\bar{y_0}}=\bar{T_0}+\bar{z},\quad\bar{T}|_{\bar{y}=\bar{y_1}}=\bar{T_0}+\bar{z}, \\
&\bar{T}|_{\bar{z}=\bar{z_0}}=\bar{T_0}+\bar{z_0},\quad\bar{T}|_{\bar{z}=\bar{z_1}}=\bar{T_0}+\bar{z_1},
 \end{aligned}
\end{eqnarray}
The initial conditions are:
\begin{eqnarray}
\begin{aligned}
&\bar{u}|_{\bar{t}=0}=\bar{v}|_{\bar{t}=0}=\bar{w}|_{\bar{t}=0}=0,\\
&\bar{T}|_{\bar{t}=0}=\bar{T_0}+\bar{z}. 
 \end{aligned}
\end{eqnarray}
In the following, symbols without overbars will be adopted to denote non-dimensional values.

\section{Numerical models and parameters}
\begin{figure}
\begin{minipage}[t]{\linewidth}
\scalebox{0.6}[0.6]{\includegraphics[width=\linewidth]{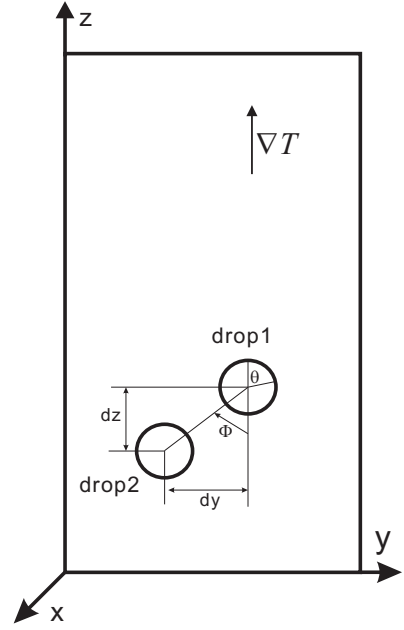}}
\caption{\label{fig:Model} The Sketch of two drops in the thermocapillary migration. $\Phi$ is the angle between the temperature gradient and the center line of two drops.}
\end{minipage}
\end{figure}

Fig. \ref{fig:Model} indicates initial positions of two drops. The horizontal and vertical
distances between two drops are $dy$ and $dz$, respectively. The traditional definition of the
non-dimensional distances are defined as $S_y=dy/2$ and $S_z=dz/2$ \cite{Ken:}, and $S_{y0}$
and $S_{z0}$ denote the initial values of $S_y$ and $S_z$. $\theta$ indicates the point on drop
interface in the $x=0$ plane: $\theta=0$ is the front stagnation and $\theta=\pi$ or $\theta=-\pi$
the rear stagnation. Points in the clockwise direction from front stagnation are denoted with
$\theta>0$, otherwise, $\theta<0$. In the full three dimensional model, the computation zone is set
to be $6\times9\times24$ on a grid of $60\times90\times240$. The time steps are $10^{-6}$ for
$Re=Ma=10^{-3}$, and $10^{-3}$ for other $Re$\&$Ma$ values. To save the computing time, the
axisymmetric model is adopted in the cases of $\Phi=0$\footnote{It has been realized that some
nonaxisymmetrical behaviors might arise for the isolated drop when the full three-dimensional
simulating domain is fairly small\cite{Brady:}. In this paper, we adopt a fairly large domain,
and assume the nonaxisymmetrical effect can be neglected for moderate parameters.}. In axisymmetric
simulations, drop1 in hotter region will be called the leading drop and drop2 in colder region the
trailing drop, and $S$ and $S_{0}$ are adopted to replace $S_{z}$ and $S_{z0}$ in the
three-dimensional model. The computing domain is $6\times24$ with the resolution of $128\times512$.
The time steps are $5\times10^{-7}$ for the simulations of $Re=Ma=10^{-3}$ and $2\times10^{-4}$ for
all other $Re$\&$Ma$ values. In this paper, all material parameters of two drops are assumed to be
the same.

Generally speaking, the non-dimensional thermocapillary migration velocities of drops are quite
small (about 0.1), and the usually defined non-dimensioan velocity (namely
$\bar{\textbf{u}}=\textbf{u}/U_{max}$, where $U_{max}$ is the maximum velocity in the flow field
instead of $U=|\sigma_T||\nabla T_\infty|a/\mu_b$) is even lower. The influence of the Re number in
the current study is trivial and can be inferred from the results of the isolated drop. The role of
the Re number will not be discussed here (simply set $Re=1$ if not specified), and we will
concentrate on the influences of thermal convection and initial distance.

To have a clear idea of the interaction of two thermocapillary drops, it is necessary to compare it
with that of the isolated drop. In the following, the velocity of the isolated drop will be denoted
as $W_{iso}$.

\begin{figure}
\begin{minipage}[t]{\linewidth}
\scalebox{1}[1]{\includegraphics[width=0.9\linewidth]{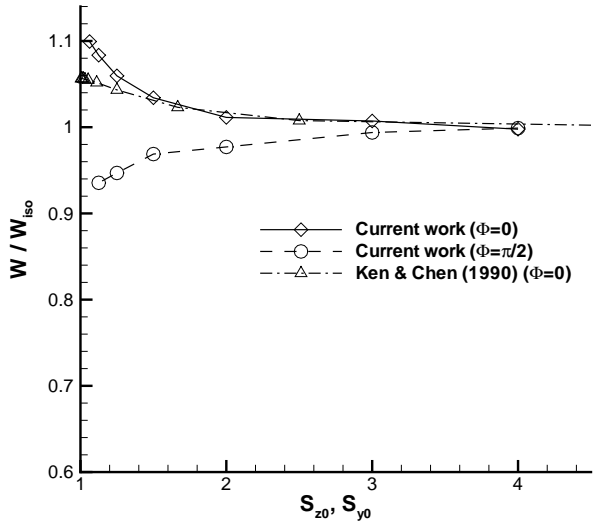}}
\caption{\label{fig:smallpara_V} The steady-state thermocapillary migration velocities of two drops with different initial distances. The horizontal ordinate is $S_{z0}$ for $\Phi=0$, and $S_{y0}$ for $\Phi=\pi/2$. Here, $Re=Ma=10^{-3}$.}
\end{minipage}
\end{figure}

\section{Thermocapillary migration of two drops with neglected inertia and thermal convection}

\begin{figure}
\begin{minipage}[t]{\linewidth}
\begin{center}
{\includegraphics[width=0.48\linewidth]{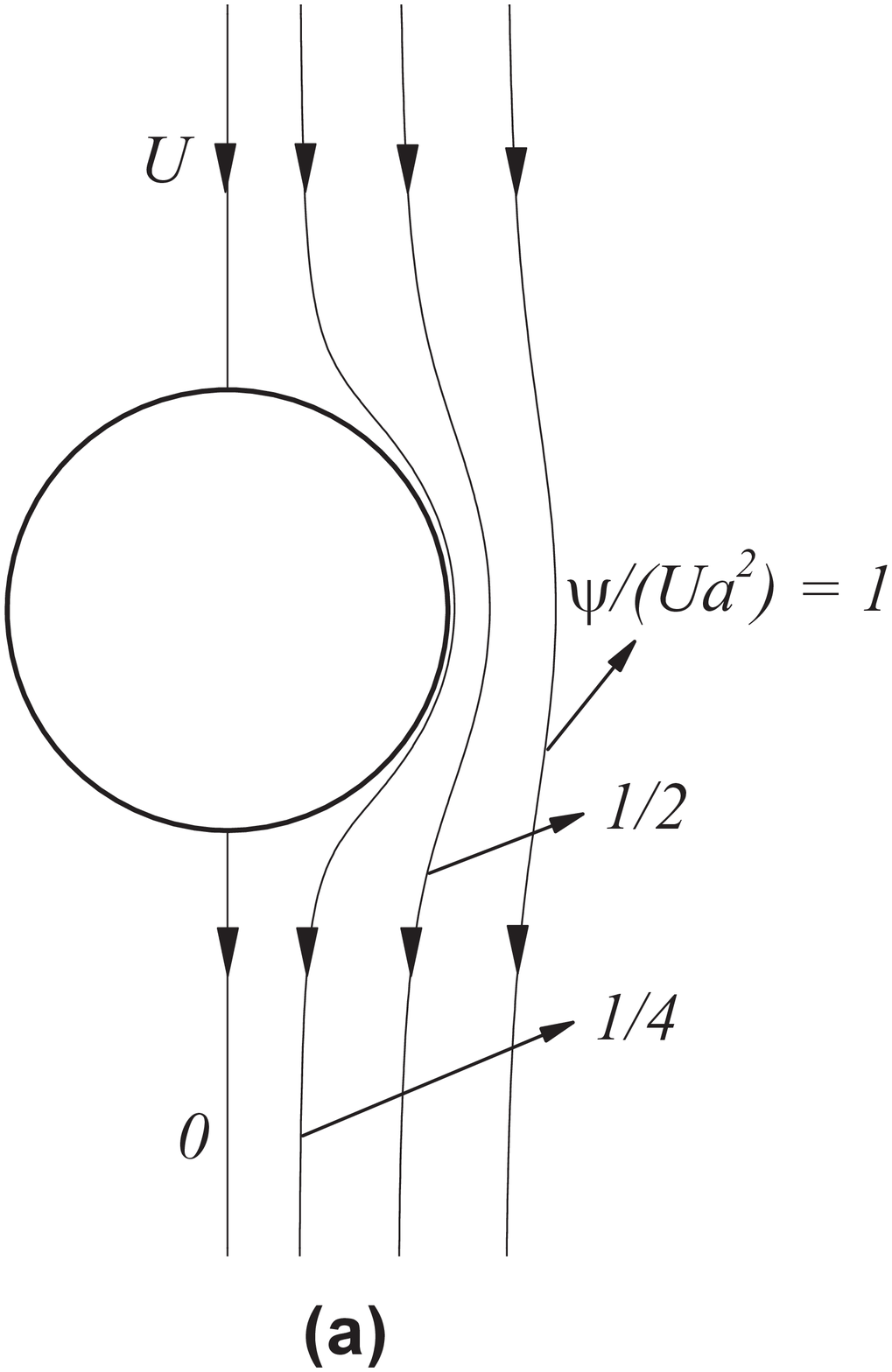}}
{\includegraphics[width=0.48\linewidth]{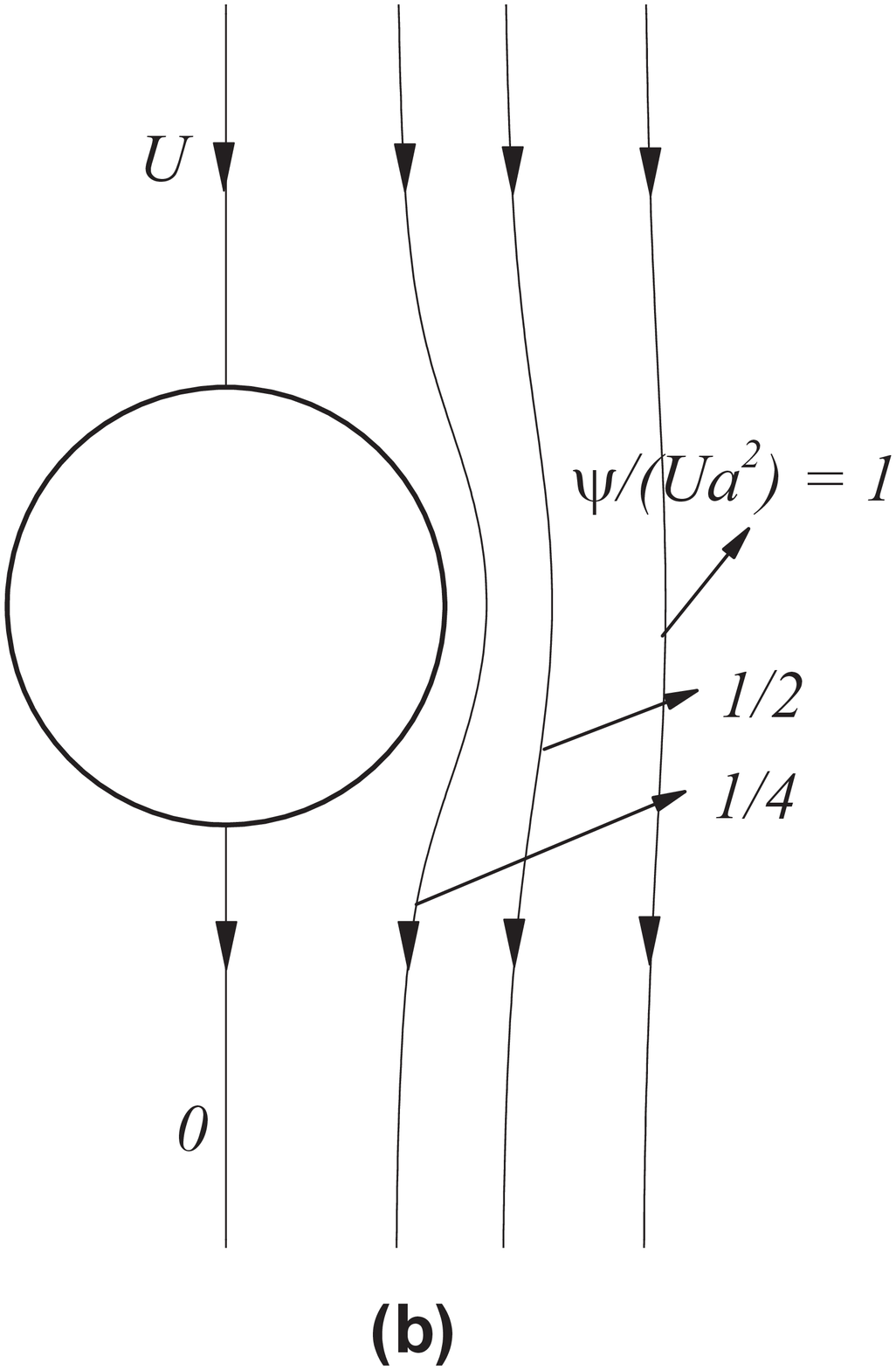}} \caption{\label{fig:stokes_poten}
Streamlines in a reference frame attached to the drop/sphere. (a) Potential flow, (b) Stokes flow.}
\end{center}
\end{minipage}
\end{figure}

In this section, small Re and Ma numbers ($10^{-3}$) are adopted in simulations to compare with
some previous analytical results. Fig. \ref{fig:smallpara_V} shows the final migration velocities
for $\Phi=0 \;  \&\;  \pi/2$. In the case of $\Phi=0$, velocities of the two drops are very close,
but faster than $W_{iso}$ (see the solid line in Fig. \ref{fig:smallpara_V}). This phenomenon is
also found in the previous analytical and numerical studies\cite{Ken:,Gao:}. In the case of
$\Phi=\pi/2$, our three dimensional simulations show that the drop velocities in $y$ or $x$
direction are neglectable. The $z$ direction velocities (W) of both drops are still the same, but
they are slower than $W_{iso}$ (the dash line in Fig. \ref{fig:smallpara_V}). In both cases, the
increase of the initial drop distance makes the final drop velocity closer to $W_{iso}$.

\begin{figure*}
\begin{minipage}[c]{0.324\linewidth}
\scalebox{1}[1.1]{\includegraphics[width=\linewidth]{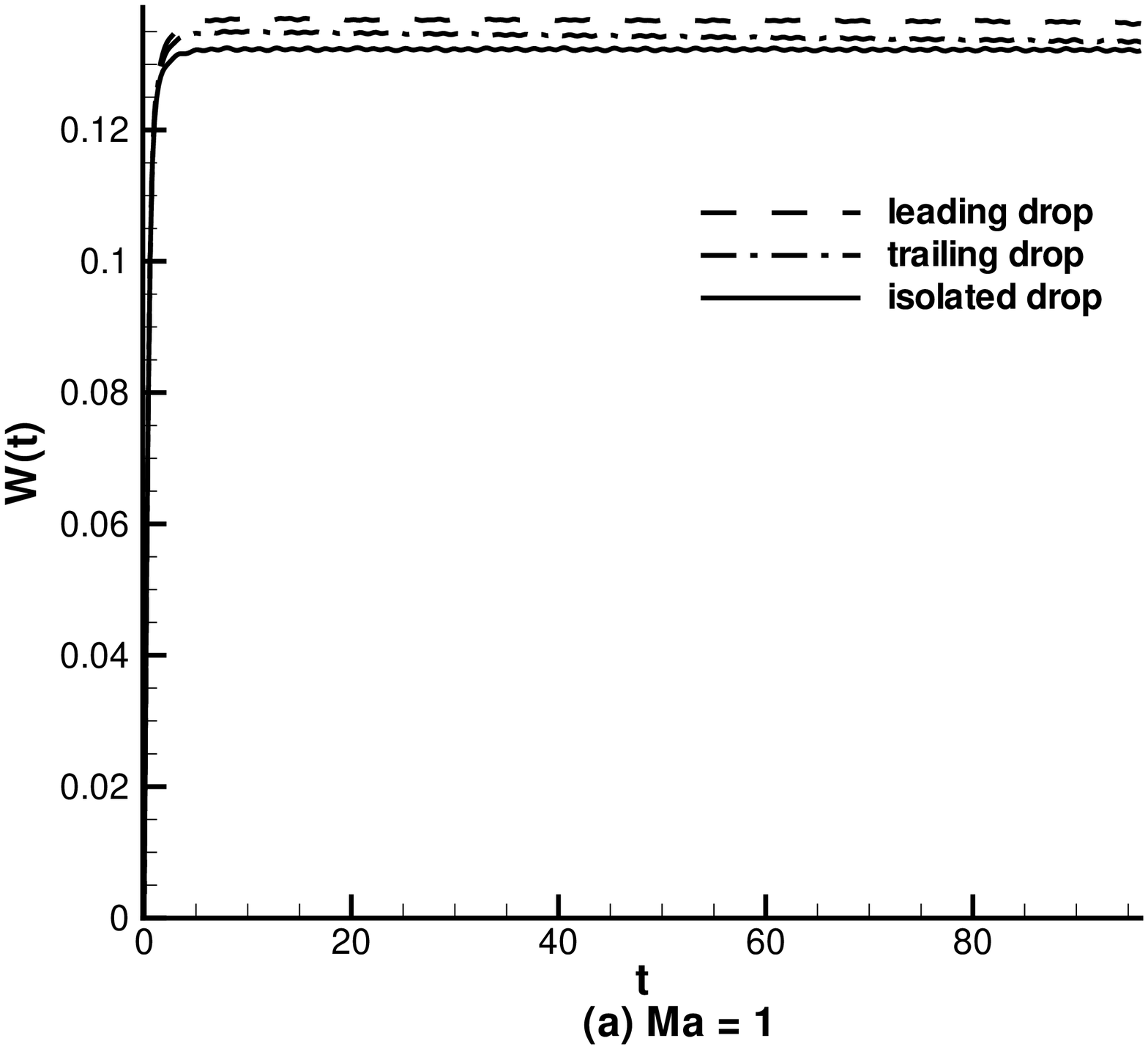}}
\end{minipage}
\begin{minipage}[c]{0.324\linewidth}
\scalebox{1}[1.1]{\includegraphics[width=\linewidth]{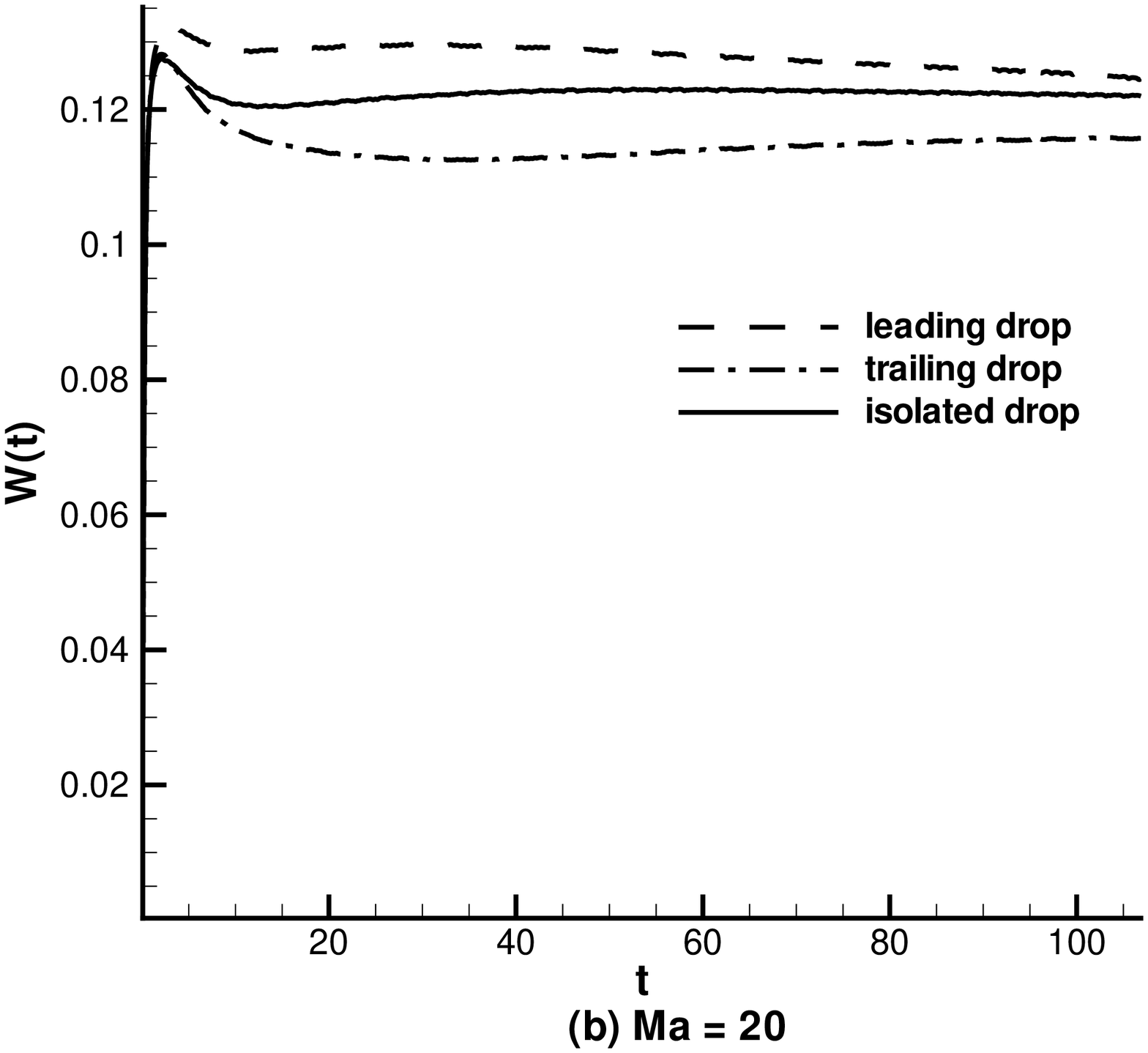}}
\end{minipage}
\begin{minipage}[c]{0.324\linewidth}
\scalebox{1}[1.1]{\includegraphics[width=\linewidth]{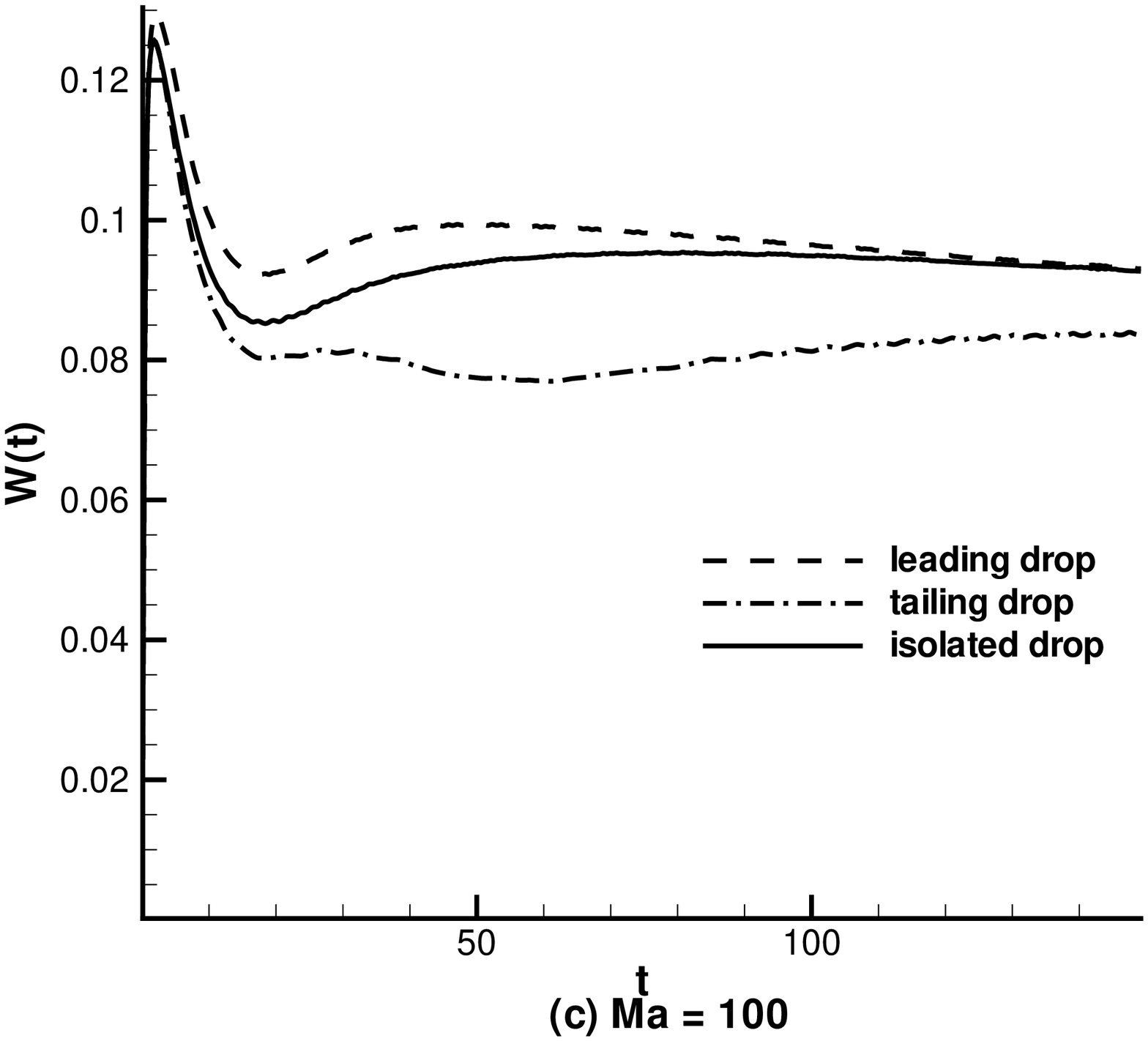}}
\end{minipage}
\caption{\label{fig:varMa_V} The migration velocities of the leading drop, the trailing drop, and the isolated drop with $Re=1$ and $S_0=1.5$.}
\end{figure*}

Similarly, there are lots of researches on the interactions between two rigid
spheres\cite{Hick:, Herman:,Kaneda:}. Using previous analytical linear results, we compare the
difference between drops and rigid spheres. Assume the rigid sphere/drop has a constant velocity U,
and the original point is in the center of the rigid sphere/drop. In the spherical coordinate $(r,
\theta, \phi)$, the potential flow, which describes the thermocapillary motion of the drop, is
written as\cite{Batchelor:}:
\begin{eqnarray}
&& v_r=U \frac{a^3}{r^3}\cos \theta,\ \ v_{\theta}=-\frac{1}{2}U
\frac{a^3}{r^3}\sin \theta.\label{eqn2}
\end{eqnarray}
Here, $\theta=0$ is the motion direction. The motion of the rigid sphere is described by the Stokes flow\cite{Happle:}:

\[
 v_r(r,\theta)=\frac{1}{2}U(3 \frac{a}{r}-\frac{a^3}{r^3})\cos
\theta
\]
\begin{equation}
 v_{\theta}(r,\theta)=-\frac{1}{4}U(3
\frac{a}{r}+\frac{a^3}{r^3})\sin \theta.\label{eqn6}
\end{equation}

It is clear that, with the increase of $r$, the velocities for the rigid sphere and drop are
decaying in the magnitude of $O(1/r^3)$ and $O(1/r)$, respectively. The velocity perturbation in
potential flow spreads in a relatively small region (Fig. \ref{fig:stokes_poten}(a)), and has
obvious directionality because the liquid in the front of the moving drop will be supplied to the
back of the drop. For Stokes flow, the velocity perturbation spreads in a fairly large region, and
the surrounding liquid tries to move with the sphere(Fig. \ref{fig:stokes_poten}(b)).

The drag coefficients of both rigid spheres are always lower than that of the isolated sphere\cite{Stimson:, Goldman:}, which means two spheres will move faster than an isolated one.  When the nonlinear effect is strong, the main interest in the interaction between rigid spheres is drag coefficients changed by the wake flow behind the leading body. However, in the thermocapillary study, the disturbed temperature field is the most important, and we will concentrate on it in the following section.

\section{Thermocapillary migration of two drops with finite inertia and thermal convection}

\subsection{Influence of thermal convection for the cases of $\Phi=0$}

Firstly, we study the influence of the Ma number on the interaction between two droplets in
thermocapillary motion with $S_0=1.5$ and $Re=1$.

In the case of $Ma=1$, the leading drop is faster than the trailing drop, but both of them move
faster than the isolated drop (Fig. \ref{fig:varMa_V}(a)). This is similar to what we have
discussed in the last section. For $Ma=20$, the leading drop is faster than the isolated drop
throughout the simulation, while the speed of the trailing drop is up to 8\% lower than $W_{iso}$
(Fig. \ref{fig:varMa_V}(b)). For $Ma=100$, the trailing drop is even slower, and its velocity is up
to 20\% lower than $W_{iso}$ (Fig. \ref{fig:varMa_V}(c)).

\begin{figure*}
\begin{minipage}[t]{0.9\linewidth}
{\includegraphics[width=0.27\linewidth]{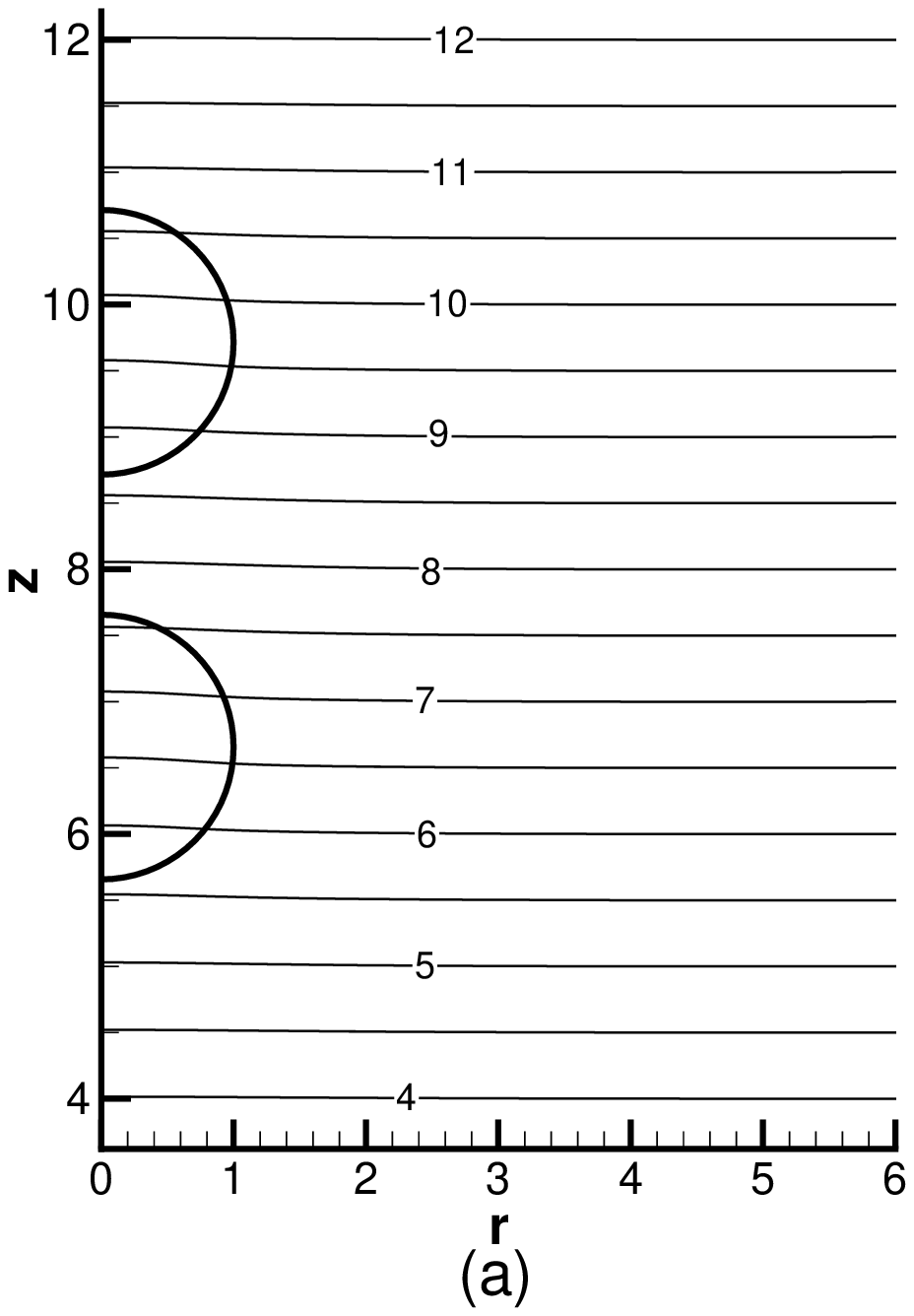}}\hspace{0.2in}
{\includegraphics[width=0.27\linewidth]{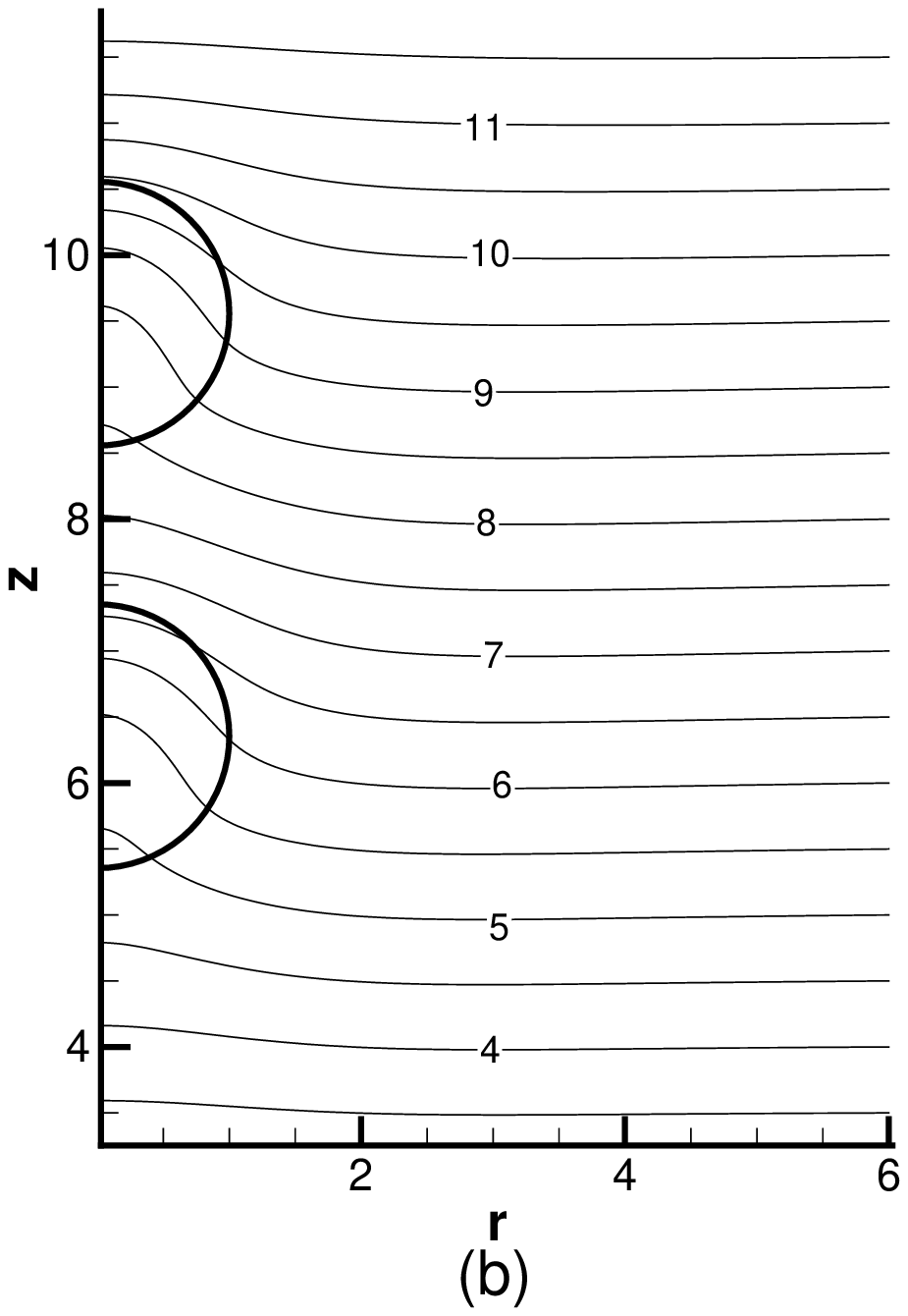}}\hspace{0.2in}
{\includegraphics[width=0.27\linewidth]{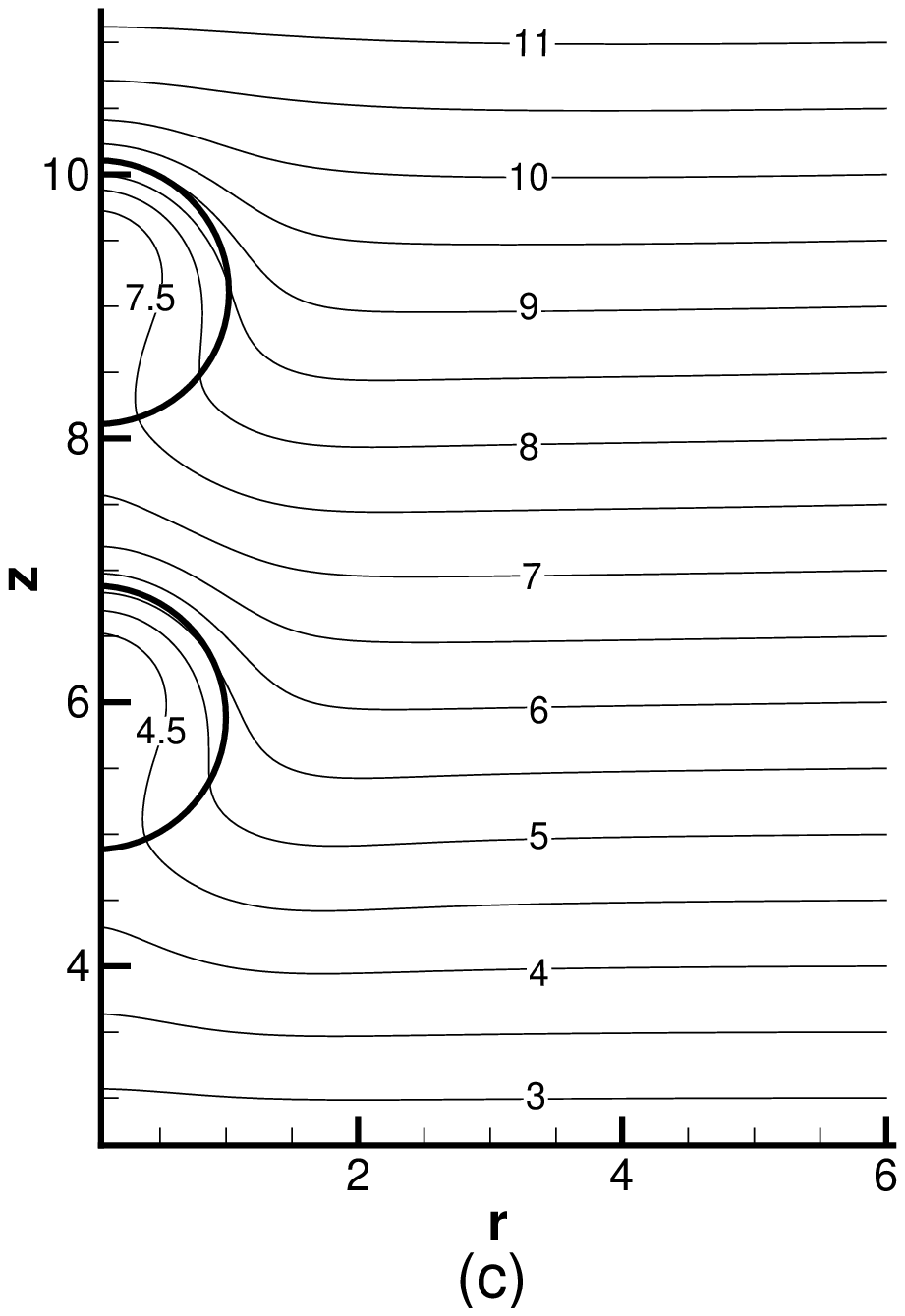}}\hspace{0.2in}
{\includegraphics[width=0.27\linewidth]{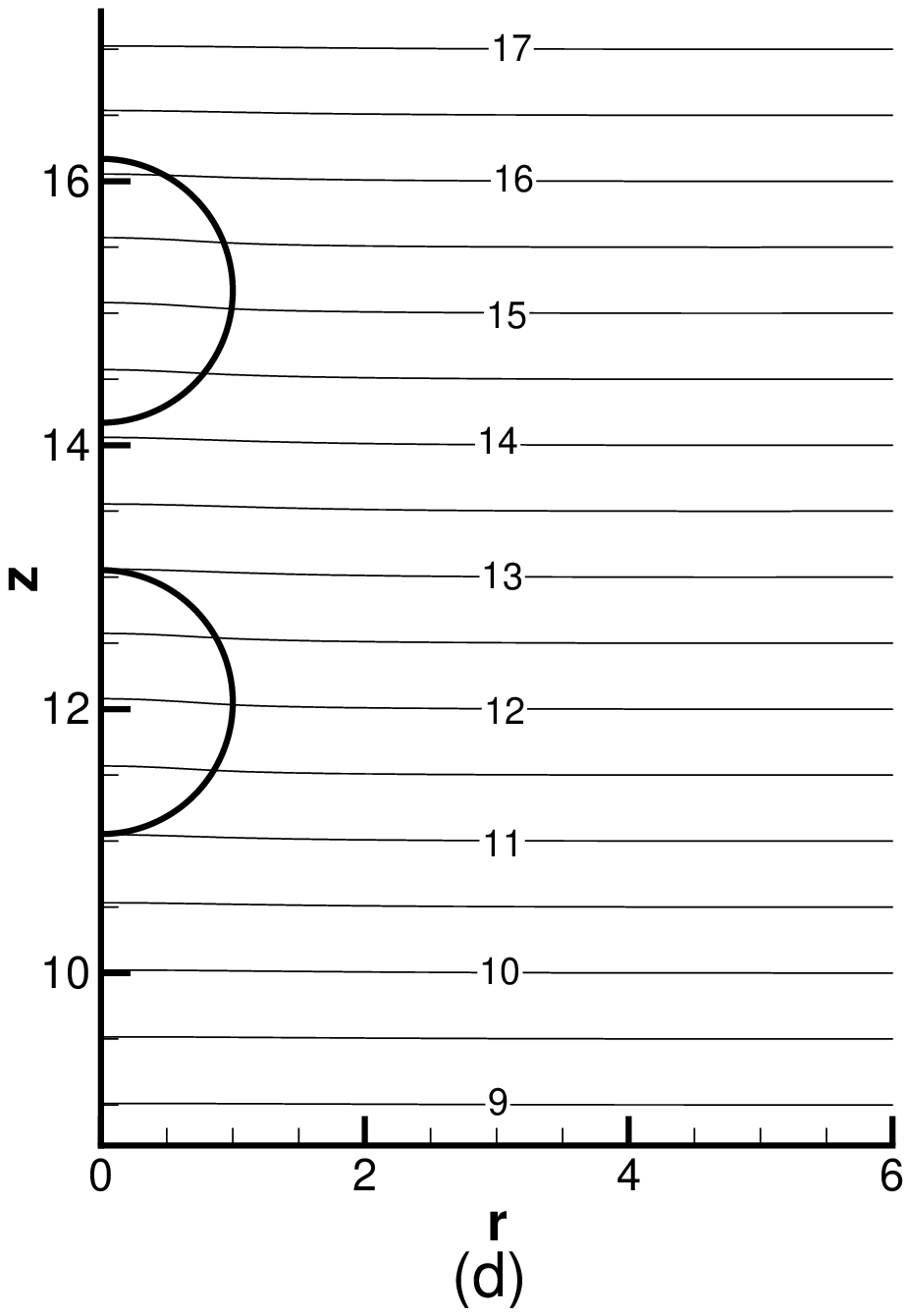}}\hspace{0.2in}
{\includegraphics[width=0.27\linewidth]{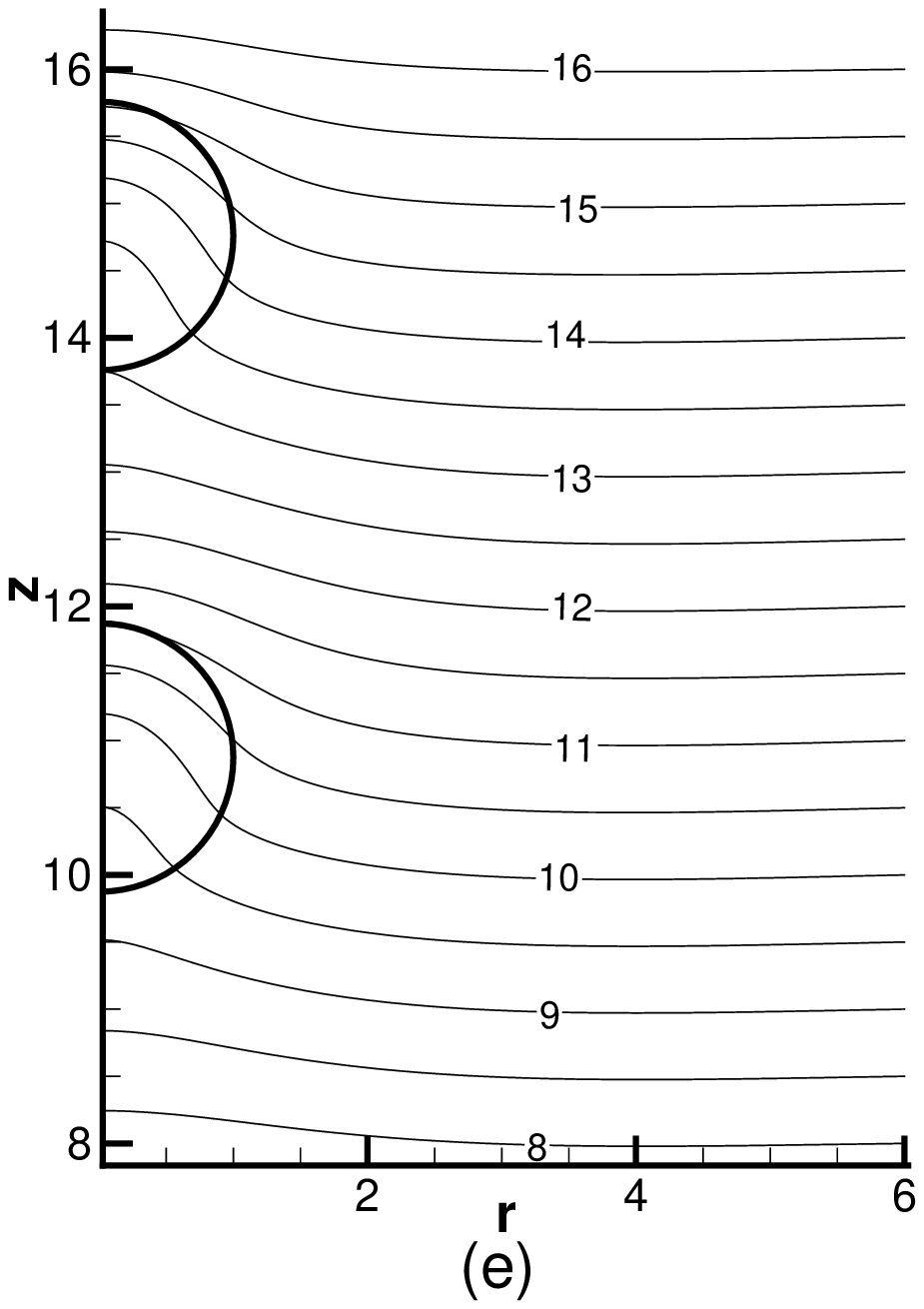}}\hspace{0.2in}
{\includegraphics[width=0.27\linewidth]{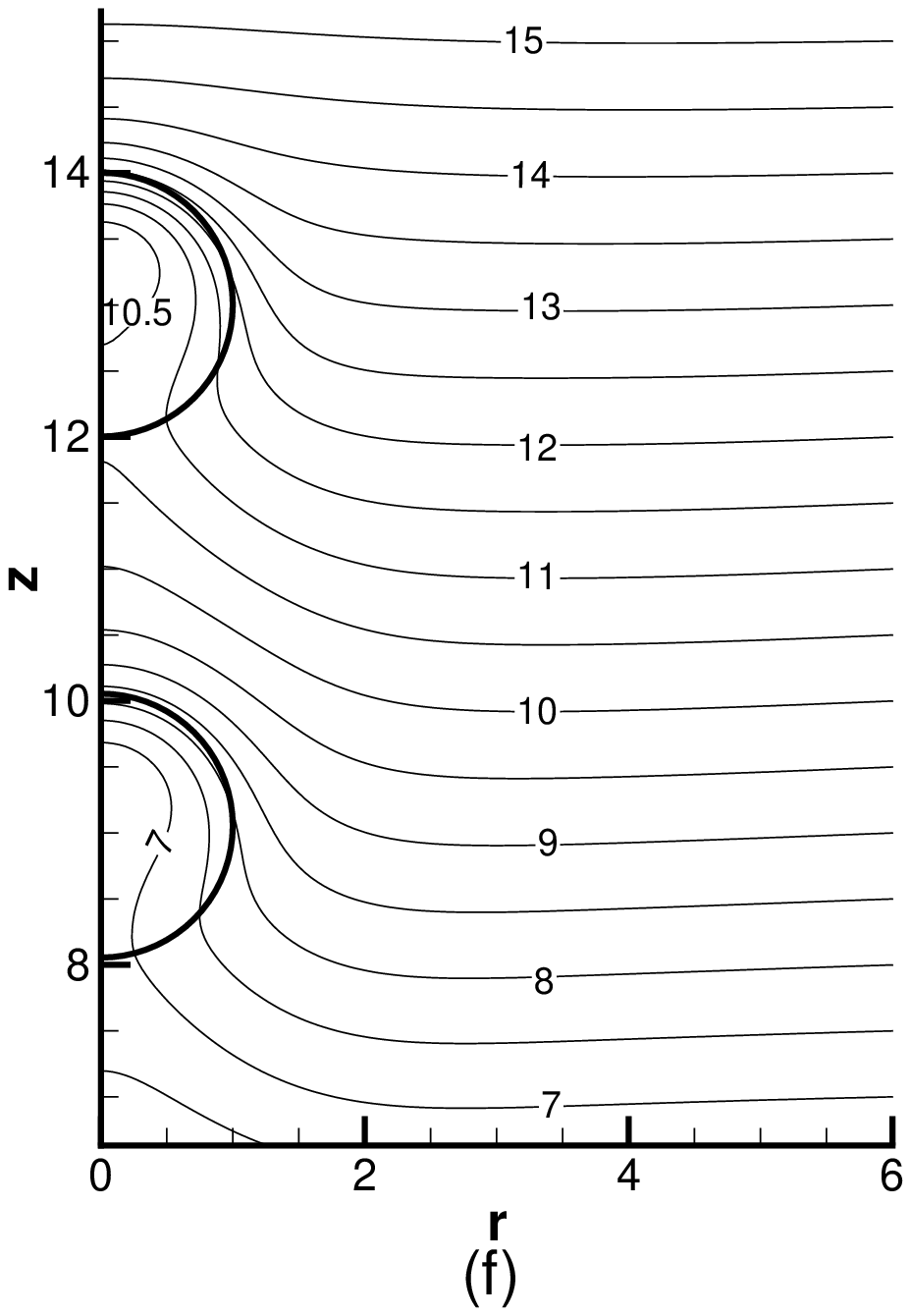}}
\end{minipage}
\caption{\label{fig:varMa_iso} The isotherms around the two droplets with $Re=1$ and $S_0=1.5$. The
first row is at $t=20$, and the second row $t=60$. The first column: $Ma=1$; the second column:
$Ma=20$; the third column: $Ma=100$.}
\end{figure*}

Figs. \ref{fig:varMa_iso} are the isotherms around the drops. When the Ma number is small, the
isotherms around the drops are almost straight and evenly spaced throughout the simulation (Figs.
\ref{fig:varMa_iso}(a)(d)). When the Ma number is increased, isotherms near $r=0$ arch to the
hotter region, and there is a closed cold zone arising in the droplet (Figs.
\ref{fig:varMa_iso}(c)(f)).  The temperature at the rear stagnation point of the leading drop is
lower than that of the isolated drop, and the temperature gradient of the trailing drop is also
reduced.

\begin{figure}
\begin{minipage}[t]{\linewidth}
\scalebox{1}[1]
{\includegraphics[width=0.9\linewidth]{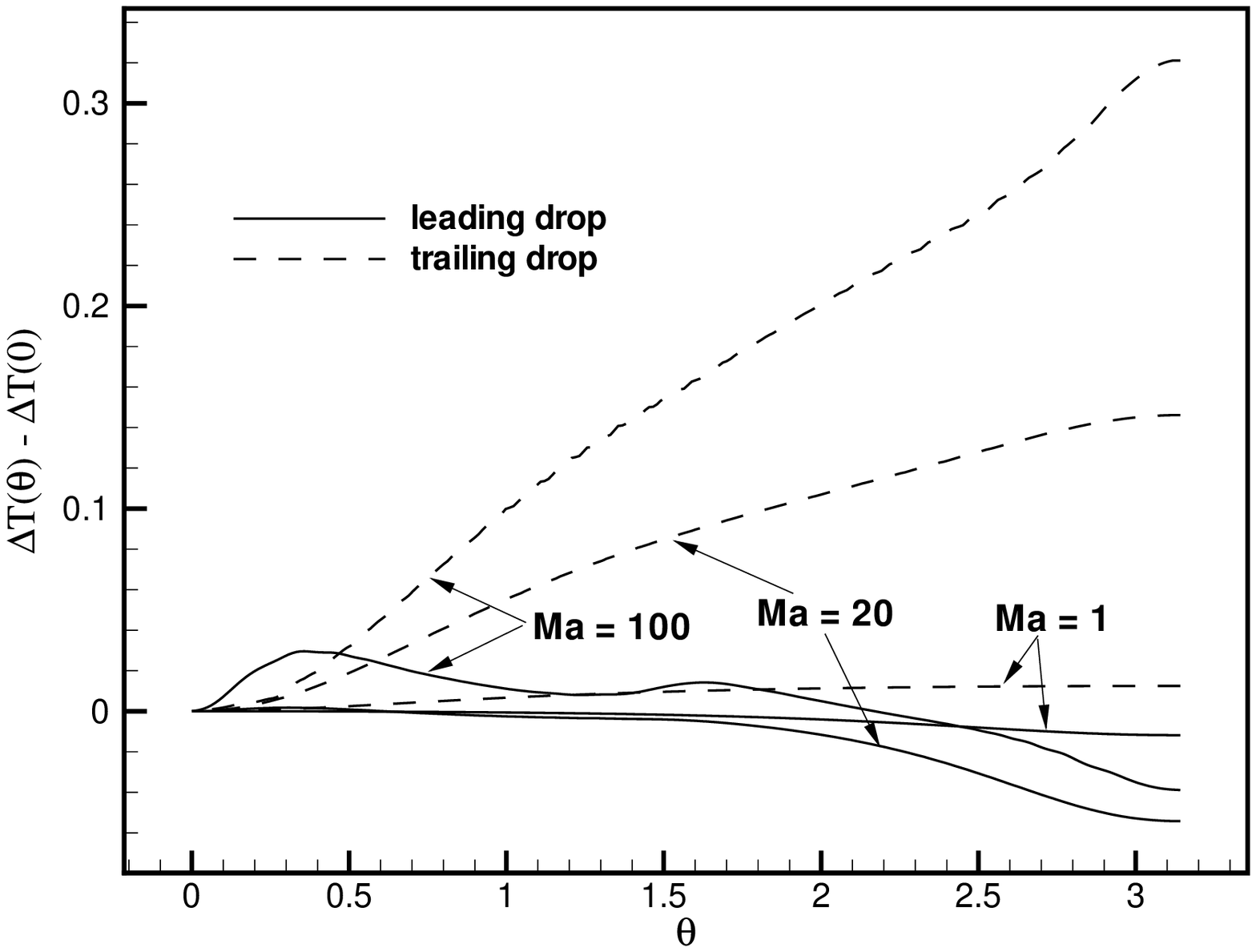}}
\caption{\label{fig:varMafaceT} Temperature difference between the point $\theta$ on interface and the front stagnation at $t=60$. Here, $Re=1$, $S_0=1.5$, and $\Delta T(\theta)=T(\theta)-T_{iso}(\theta)$. }
\end{minipage}
\hspace{0.4in}
\begin{minipage}[t]{0.9\linewidth}
\scalebox{1}[1]
{\includegraphics[width=\linewidth]{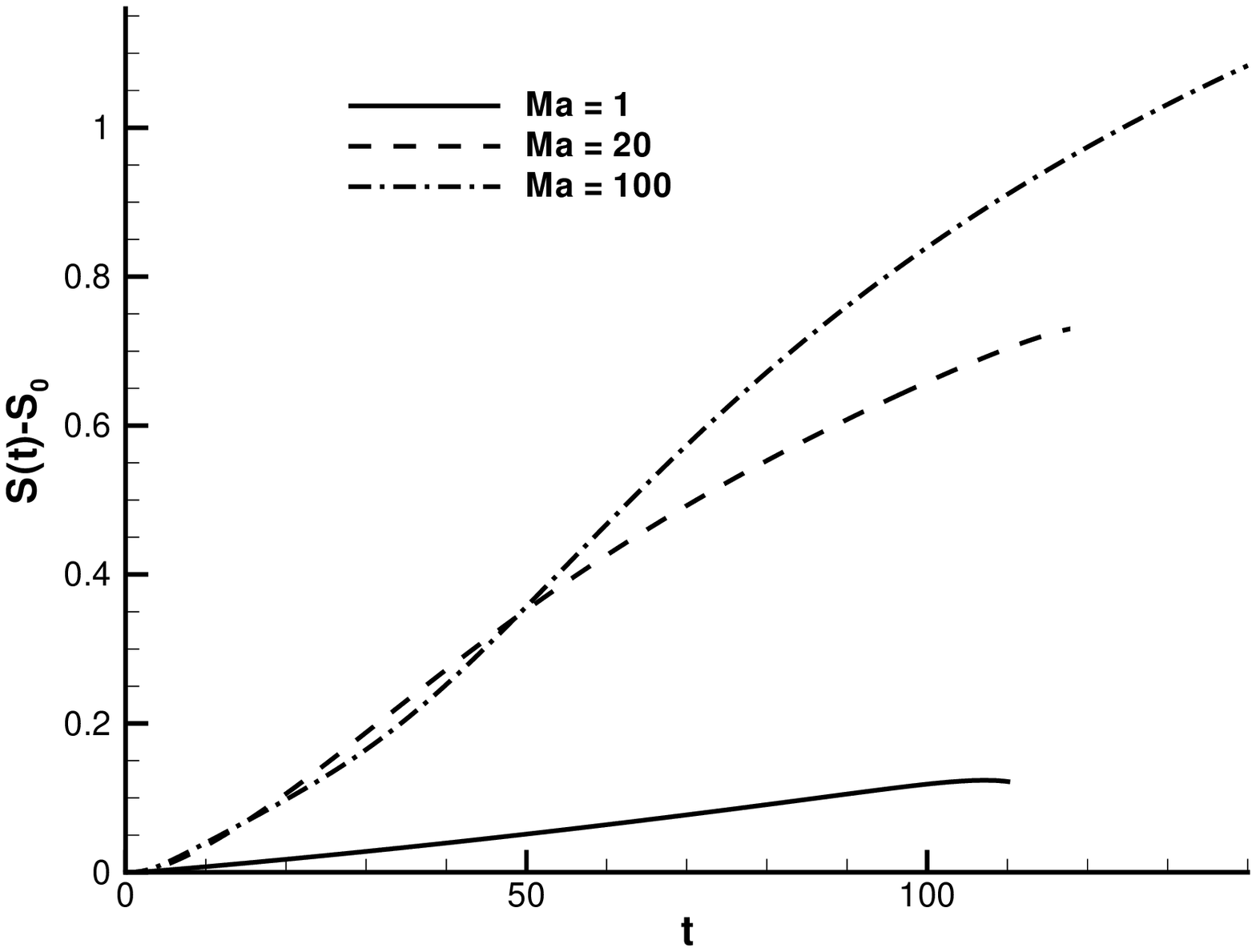}}
\caption{\label{fig:S-S0_varMa} Time evolutions of separation distances ($S-S_0$) between two drops for $Re=1$ and $S_0=1.5$.}
\end{minipage}
\end{figure}

\begin{figure}
\begin{minipage}[t]{0.9\linewidth}
\scalebox{1}[1]{\includegraphics[width=\linewidth]{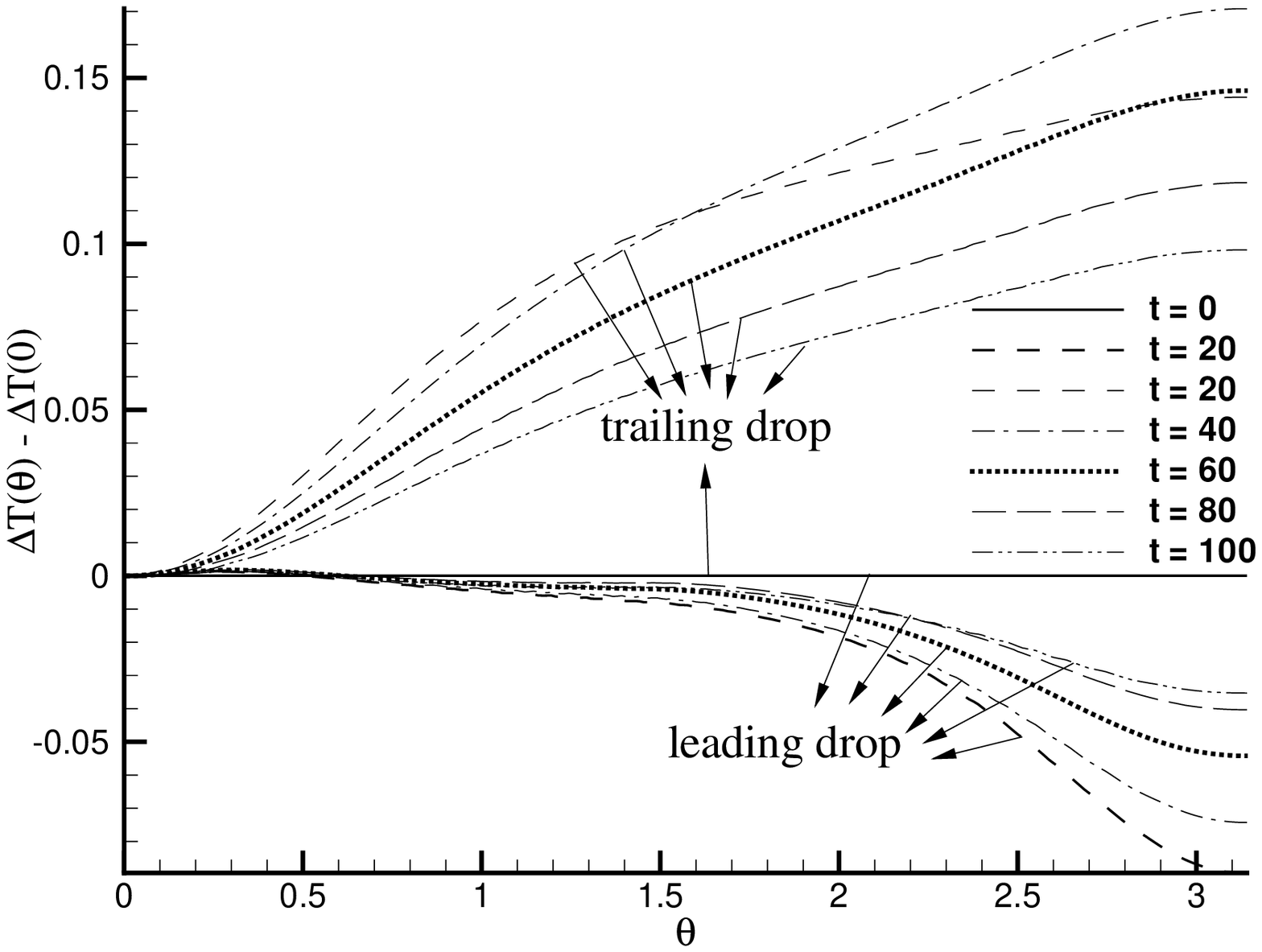}}
\caption{\label{fig:2d_Re1Ma20_faceT} The time evolution of the temperature difference for $Re=1$, $Ma=20$ and $S_0=1.5$. }
\end{minipage}
\end{figure}

Fig. \ref{fig:varMafaceT} shows the temperature difference between the point on drop surface and
the front stagnation (to get a clear idea of this difference, the corresponding value of the
isolated drop is subtracted in Figs. \ref{fig:varMafaceT}, \ref{fig:2d_Re1Ma20_faceT} and
\ref{fig:3d_Re1Ma20_t=60_surfaceT}). It can be seen that the temperature difference between the
front and rear stagnation points of the leading drop (solid line) is larger than that of the
isolated drop; on the contrary, the difference for the trailing drop (dashed line) is smaller than
that of the isolated drop, and will decrease with increasing Ma numbers.  Fig. \ref{fig:varMafaceT}
clearly shows that the influence of the thermal convection on trailing drops is much stronger than
that on leading drops.

When the heat convection is stronger, the influence on the trailing drop is also bigger. Hence, the
separated distances ($S-S_0$) increase more rapidly for larger Ma numbers (Fig.
\ref{fig:S-S0_varMa}). Note that the separating speed of $Ma=100$ before $t=60$ is lower than that
of $Ma=20$. This is because the fluctuation process of migrating speed in the beginning is longer
for larger $Ma$ number\cite{Yin:}. For example, the thermal wake left by leading drop with
$Ma=100$ is not fully developed until $t=60$ (Fig. \ref{fig:varMa_iso}(c)(f)), and the velocity
difference between leading and trailing drops is not so large.

Temperature differences for $Ma=20$ at various moments are shown in Fig.
\ref{fig:2d_Re1Ma20_faceT}. At $t=20$, the temperature difference of the leading drop becomes
bigger than that of isolated drop, and starts to become smaller afterwards. When the thermal
convection effect and the temperature disturbance caused by the leading drop are fully developed,
the temperature difference of the trailing drop reaches its minimum at $t=40$, and starts to become
larger afterwards.

\subsection{Influence of initial distance for the cases of $\Phi=0$}

Obviously, there will be no interaction between drops if they are far enough from each other, so it
is interesting to know the critical initial distance at which both droplets migrate like an
isolated one.

\begin{figure*}
\begin{minipage}[t]{0.9\linewidth}
{\includegraphics[width=0.4\linewidth]{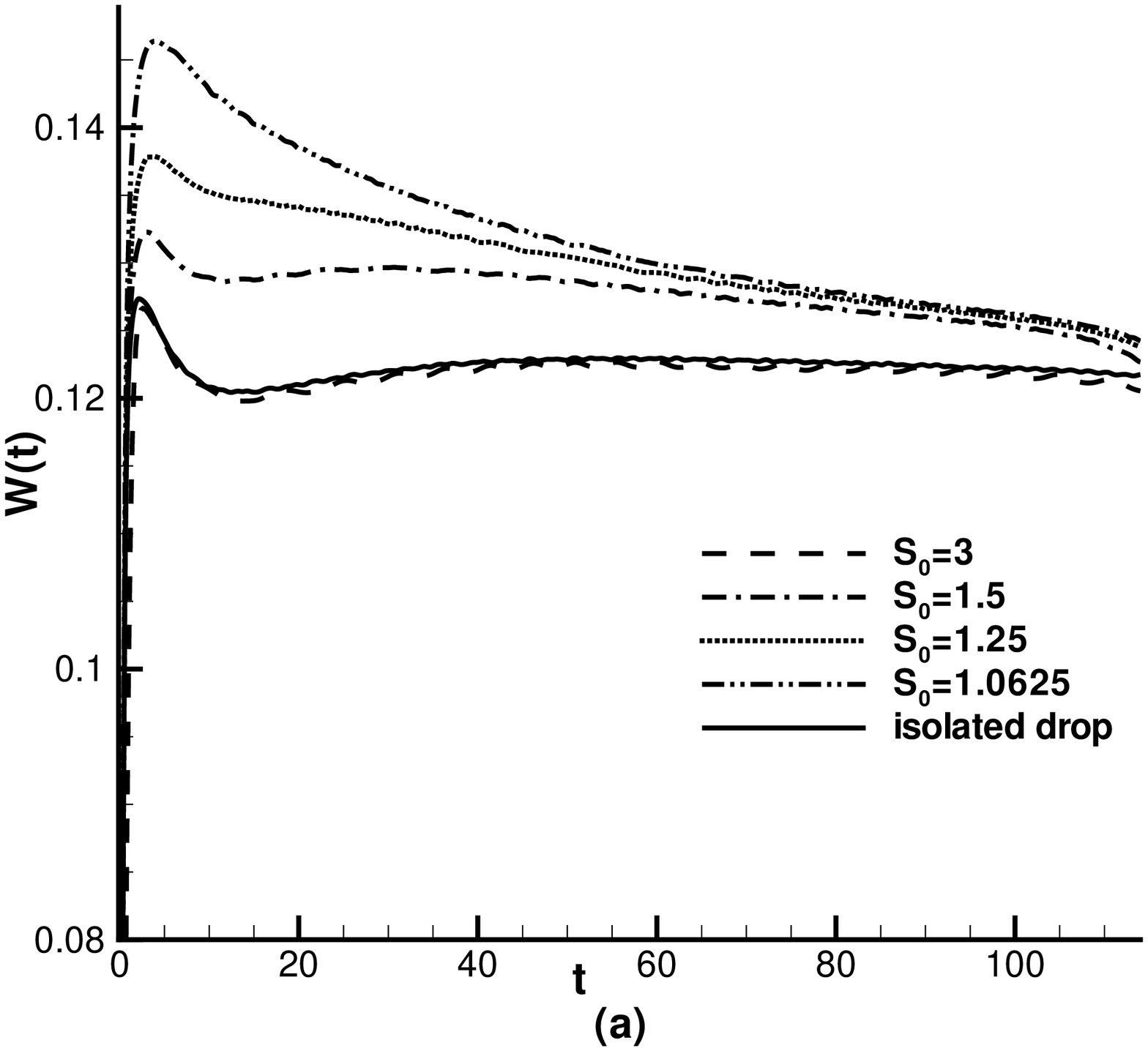}}\hspace{0.5in}
{\includegraphics[width=0.4\linewidth]{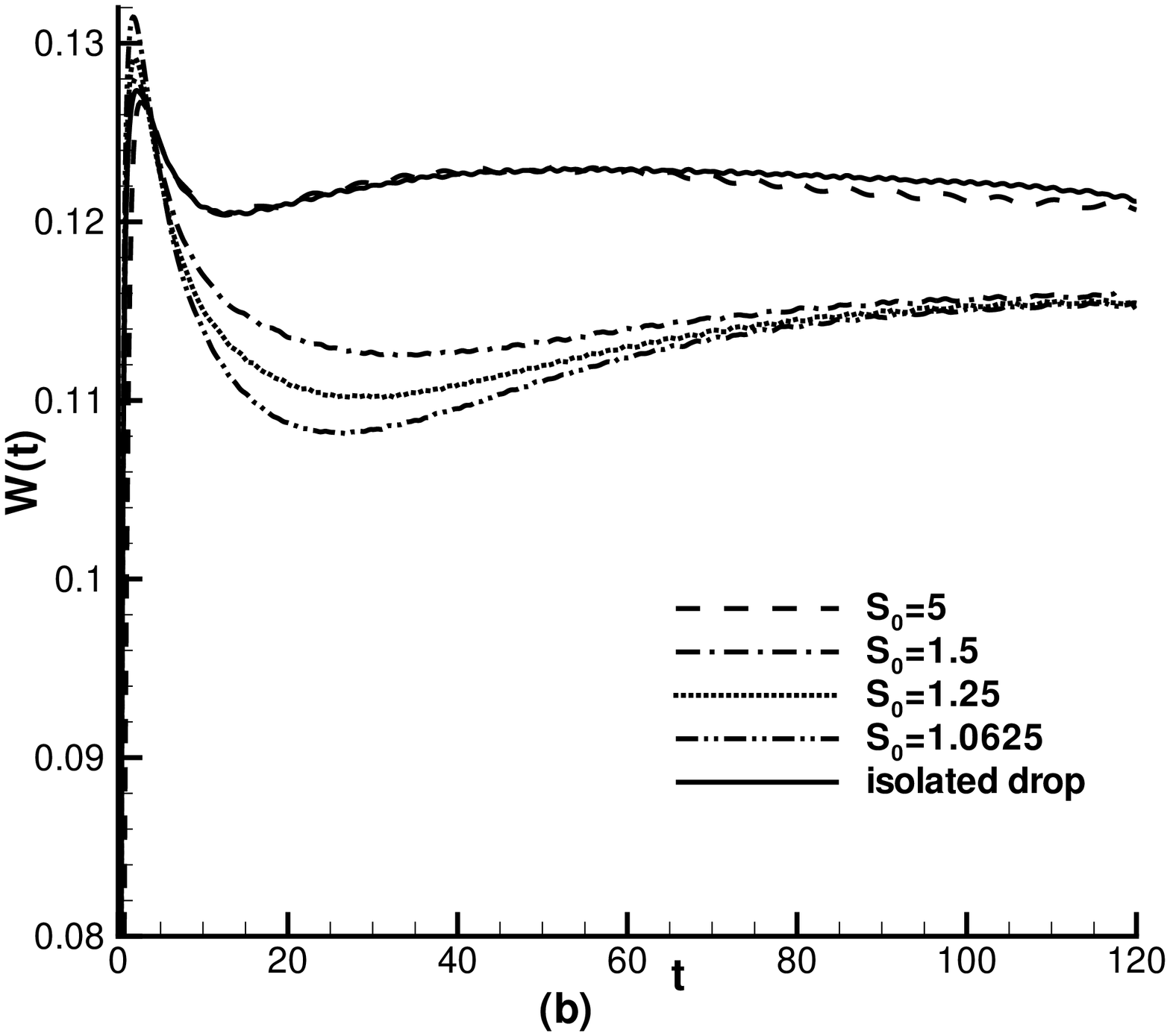}}
\end{minipage}
\caption{\label{fig:vardz_Ma20_V} Time evolutions of drop velocities for $Re=1$ and $Ma=20$ with various initial distances. (a) Leading drop, (b) trailing drop.}
\end{figure*}

Several different initial distances are tested for the case of $Re=1$ and $Ma=20$ (Figs.
\ref{fig:vardz_Ma20_V}). The curve of the leading drop with $S_0=3$ is almost identical to that of
the isolated drop (Fig. \ref{fig:vardz_Ma20_V}(a)), while the critical initial distance is $5$ for
the trailing drop (Fig. \ref{fig:vardz_Ma20_V}(b)). It seems that the thermal wake left by the
leading drop affects the trailing drop at a longer distance than the distance at which the trailing
drop interferes the leading drop.

\begin{figure*}
{\includegraphics[width=0.4\linewidth]{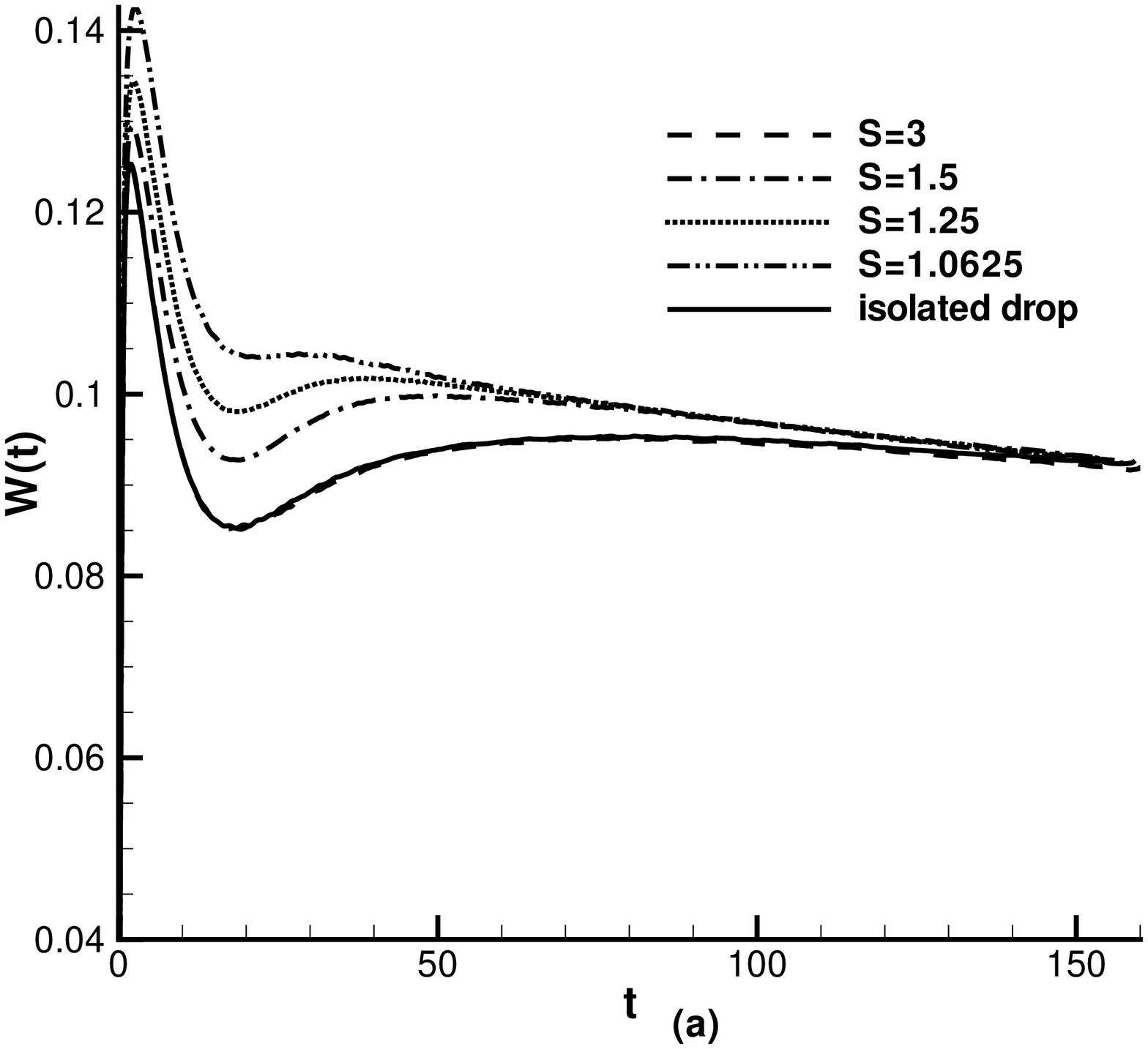}}\hspace{0.5in}
{\includegraphics[width=0.4\linewidth]{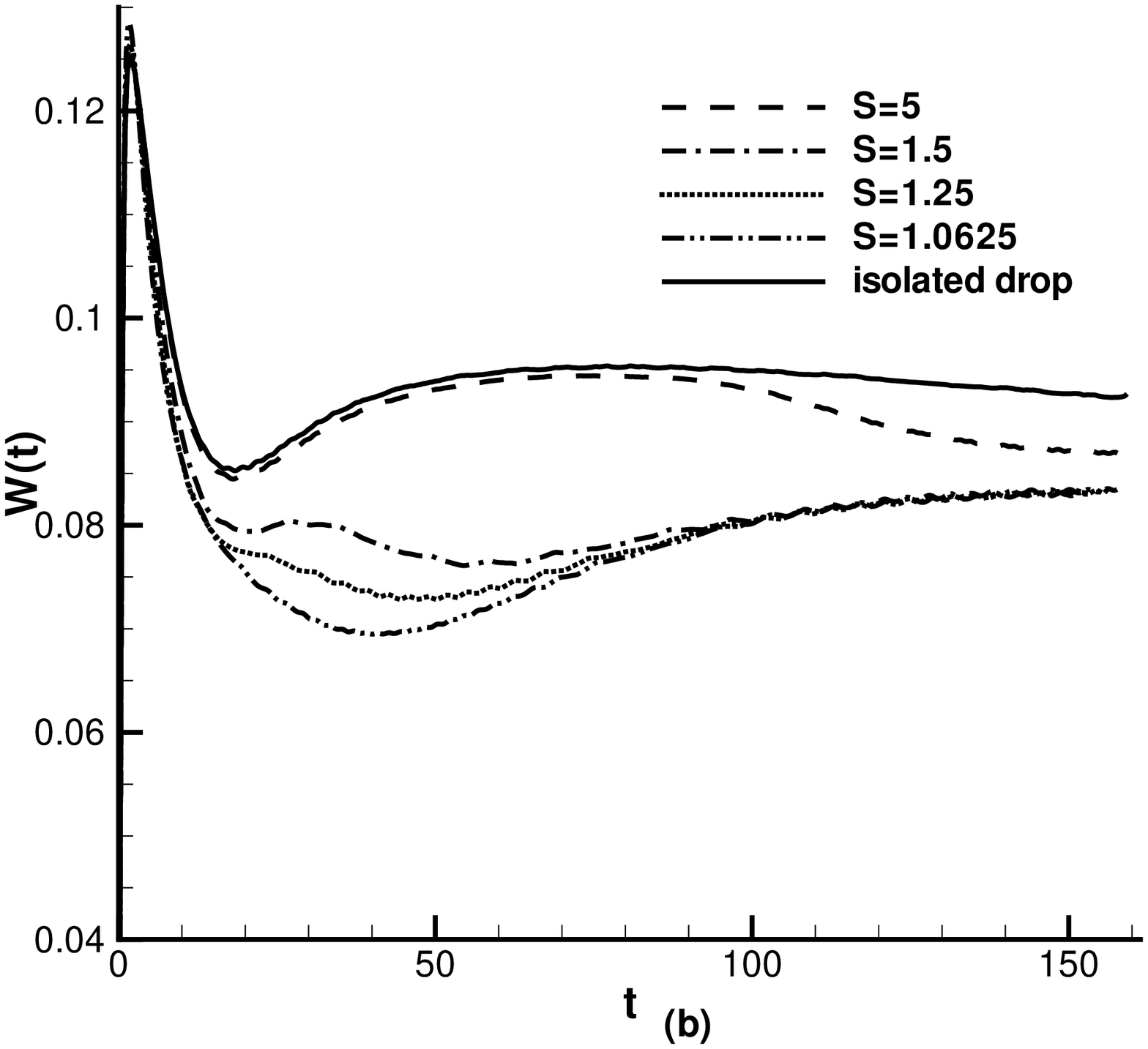}}
\caption{\label{fig:vardz_Ma100_V} $Re=1$, $Ma=100$,  time evolutions of drop velocities with various initial distances. (a) Leading drop, (b)trailing drop.}
\end{figure*}

When the $Ma$ number is increased to 100, the critical initial distance for the leading drop
remains to be $S_0=3$ (Fig. \ref{fig:vardz_Ma100_V}(a)).  On the other hand, the thermal wake left
by the leading drop is much longer and the critical initial distance for the trailing drop seems
longer than $5$.

\begin{figure*}
{\includegraphics[width=0.4\linewidth]{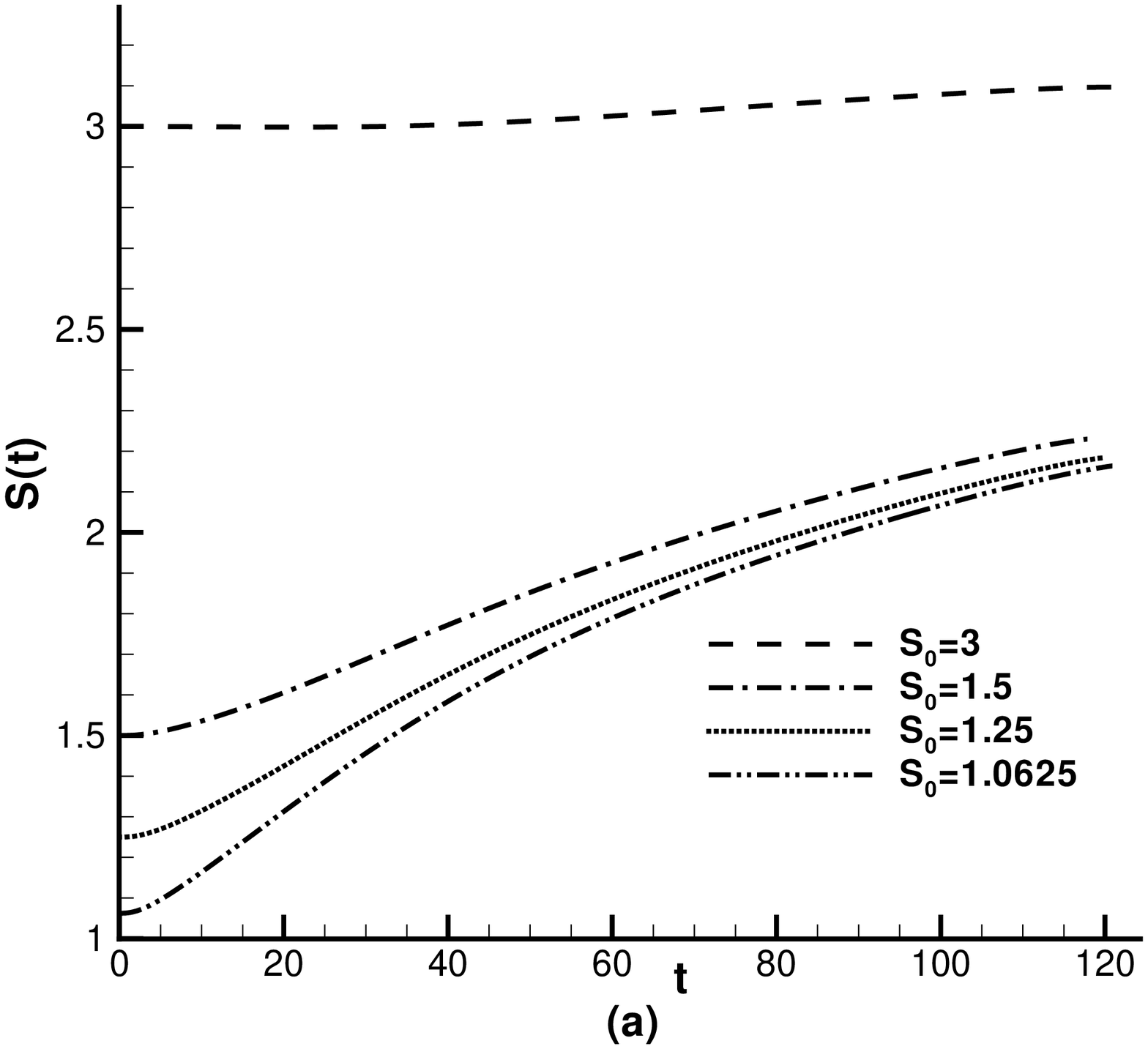}}\hspace{0.5in}
{\includegraphics[width=0.4\linewidth]{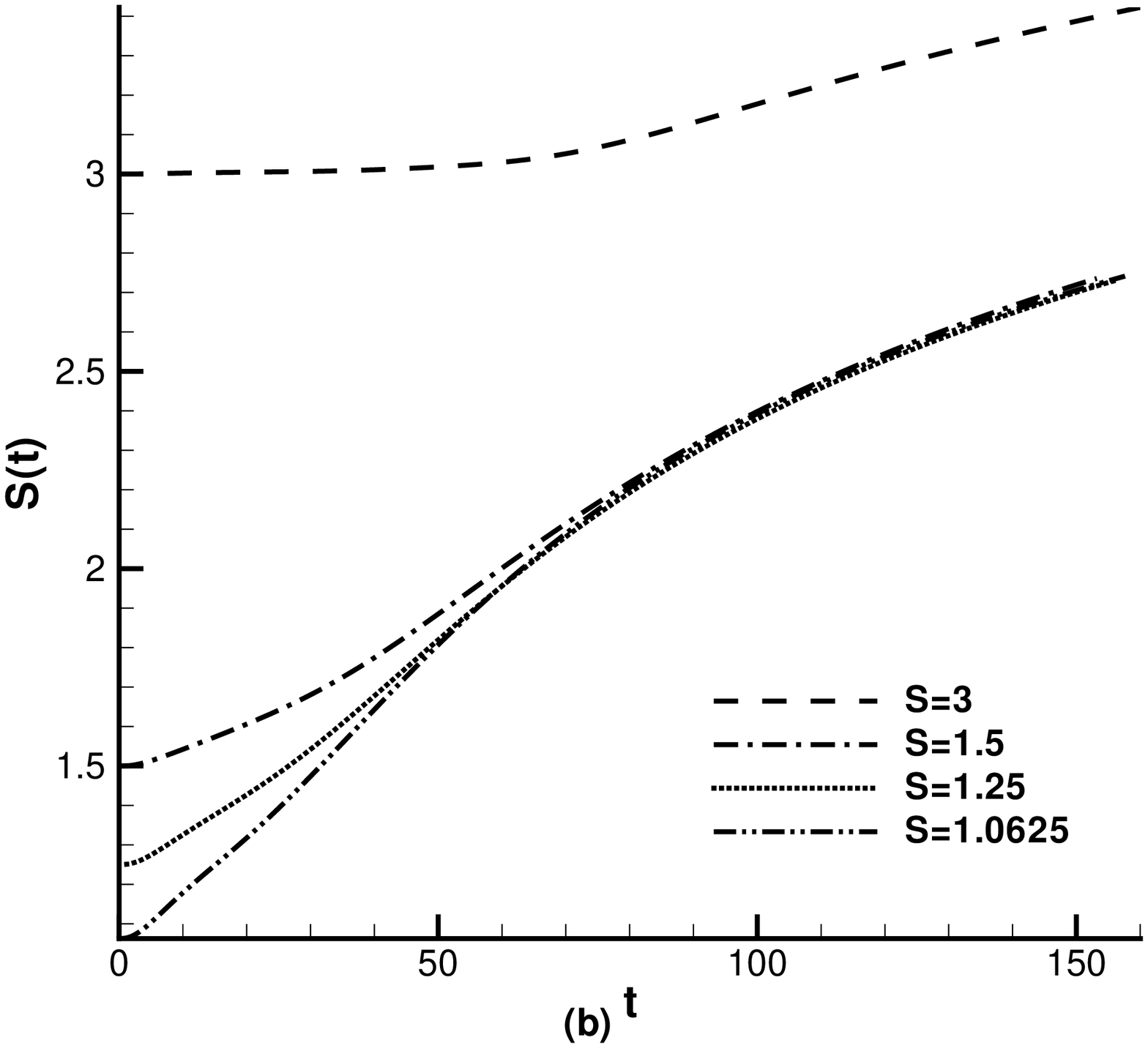}}
\caption{\label{fig:vardz_dz} Time evolutions of the separating distances between drops. (a) $Re=1$, $Ma=20$, (b) $Re=1$, $Ma=100$.}
\end{figure*}

Figs. \ref{fig:vardz_dz} shows the time evolutions of distances between two drops. It's clear that
the smaller the initial distance between the two drops, the faster they will separate from each
other.  In the later stage of simulations, it seems that the distance differences caused by various
initial distances vanish, and the final distance between drops (of course, if they are not too far
away apart) is determined by other parameters.

\subsection{Influence of thermal convection for the cases of $\Phi\neq0$}

\begin{figure*}
\begin{minipage}{0.4\linewidth}
{\includegraphics[width=\linewidth]{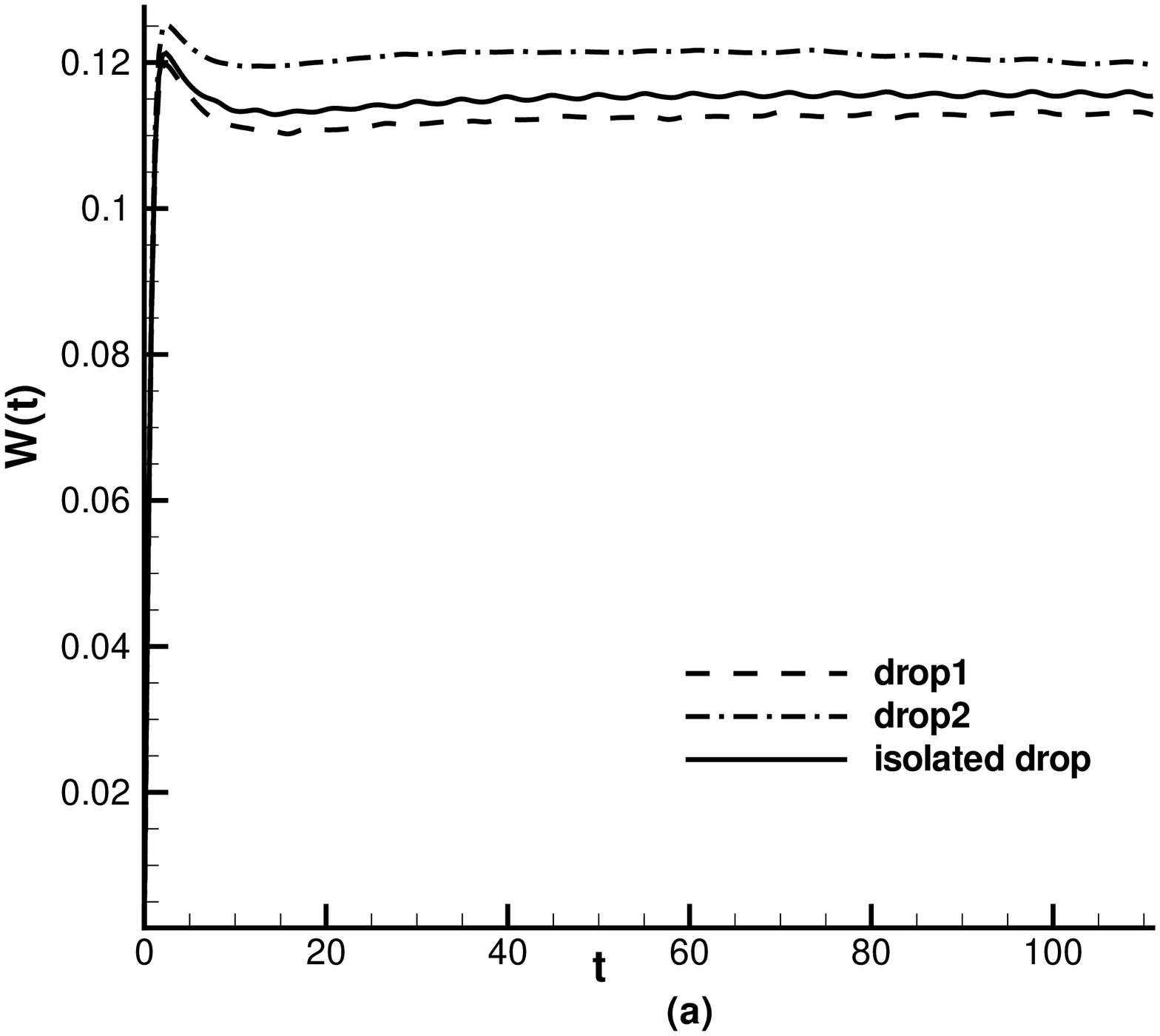}}
\end{minipage}
\hspace{0.5in}
\begin{minipage}{0.4\linewidth}
{\includegraphics[width=\linewidth]{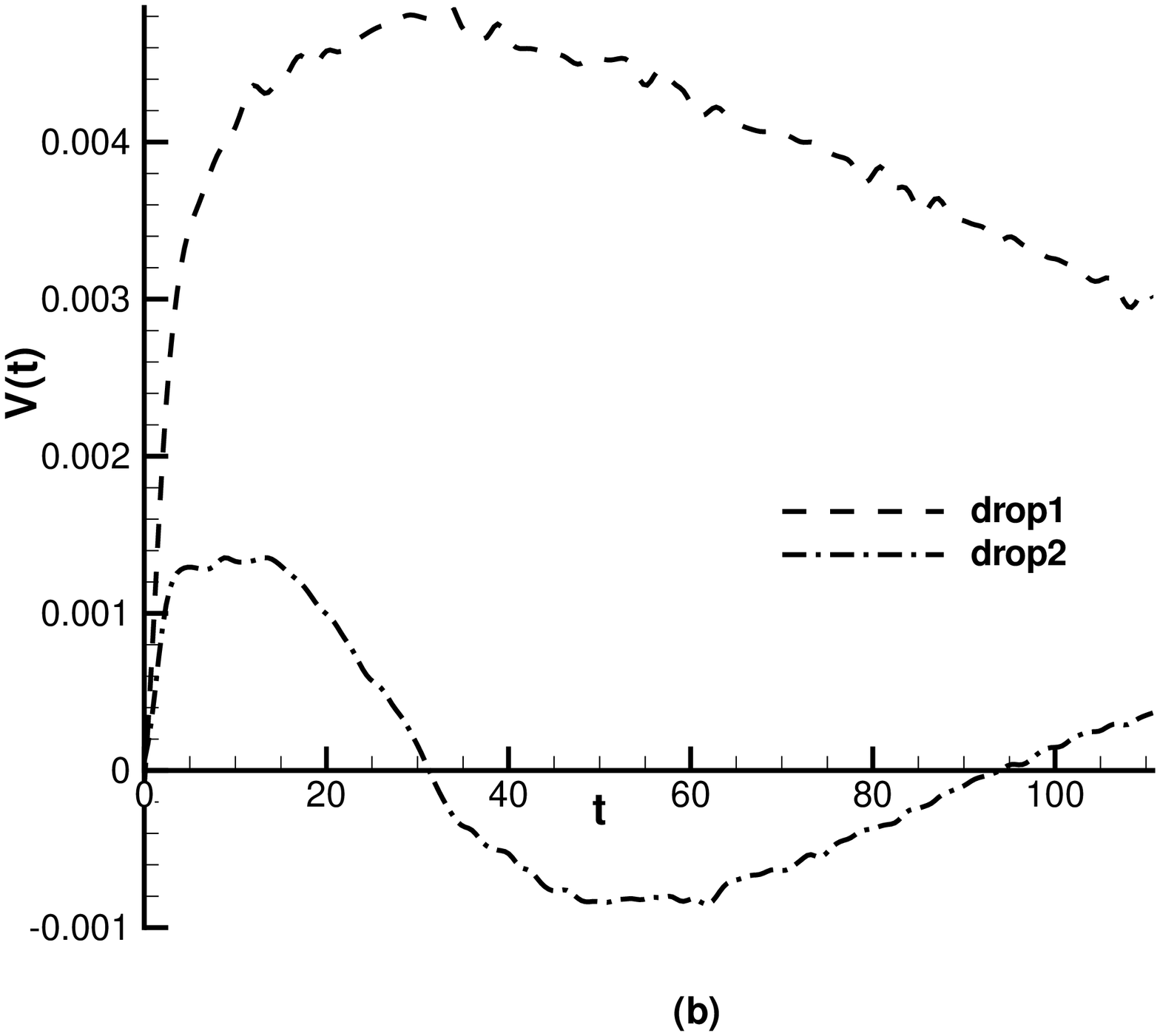}}
\end{minipage}
\caption{\label{fig:3d_Re1Ma20_V}$Re=1$, $Ma=20$ and $S_{y0}=S_{z0}=1.35$, evolution of migration velocities, (a) velocity in $z$ direction and (b) velocity in $y$ direction.}
\end{figure*}

\begin{figure}%
\scalebox{1}[1]{\includegraphics[width=0.6\linewidth]{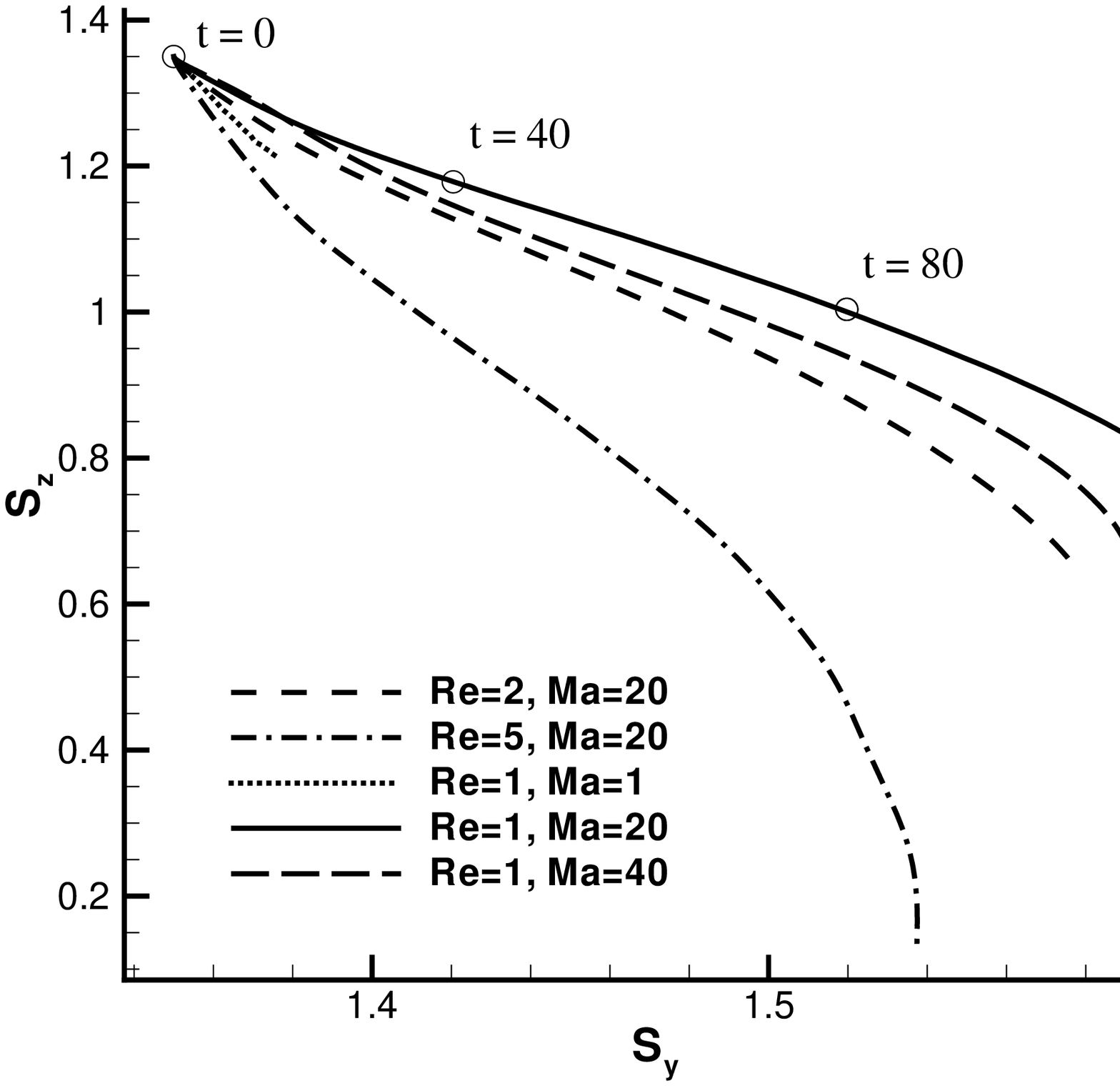}}
\caption{\label{fig:3d_varReMa_dy_dz} The time evolution of the vertical and horizontal distances between two drops with $S_{y0}=S_{z0}=1.35$.}
\end{figure}

In this subsection, the full three dimensional problem with  $Re=1$, $Ma=20$ and
$S_{y0}=S_{z0}=1.35$ is studied. In this simulation, both drops have zero velocities in the $x$
direction. In the vertical direction, the upper drop1 migrates slower than the isolated drop while
the lower drop2 moves faster than the isolated drop (Fig. \ref{fig:3d_Re1Ma20_V}(a)). As a result,
the vertical distance between the two drops is always decreasing (see the solid line in Fig.
\ref{fig:3d_varReMa_dy_dz}).

\begin{figure}
{\includegraphics[width=0.5\linewidth]{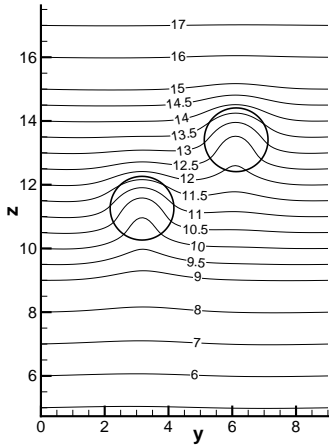}}
\caption{\label{fig:3d_Re1Ma20_t=60_iso_stream} Isotherms at $t=60$. $Re=1$, $Ma=20$ and $S_{y0}=S_{z0}=1.35$.}
\end{figure}

\begin{figure*}
{\includegraphics[width=0.4\linewidth]{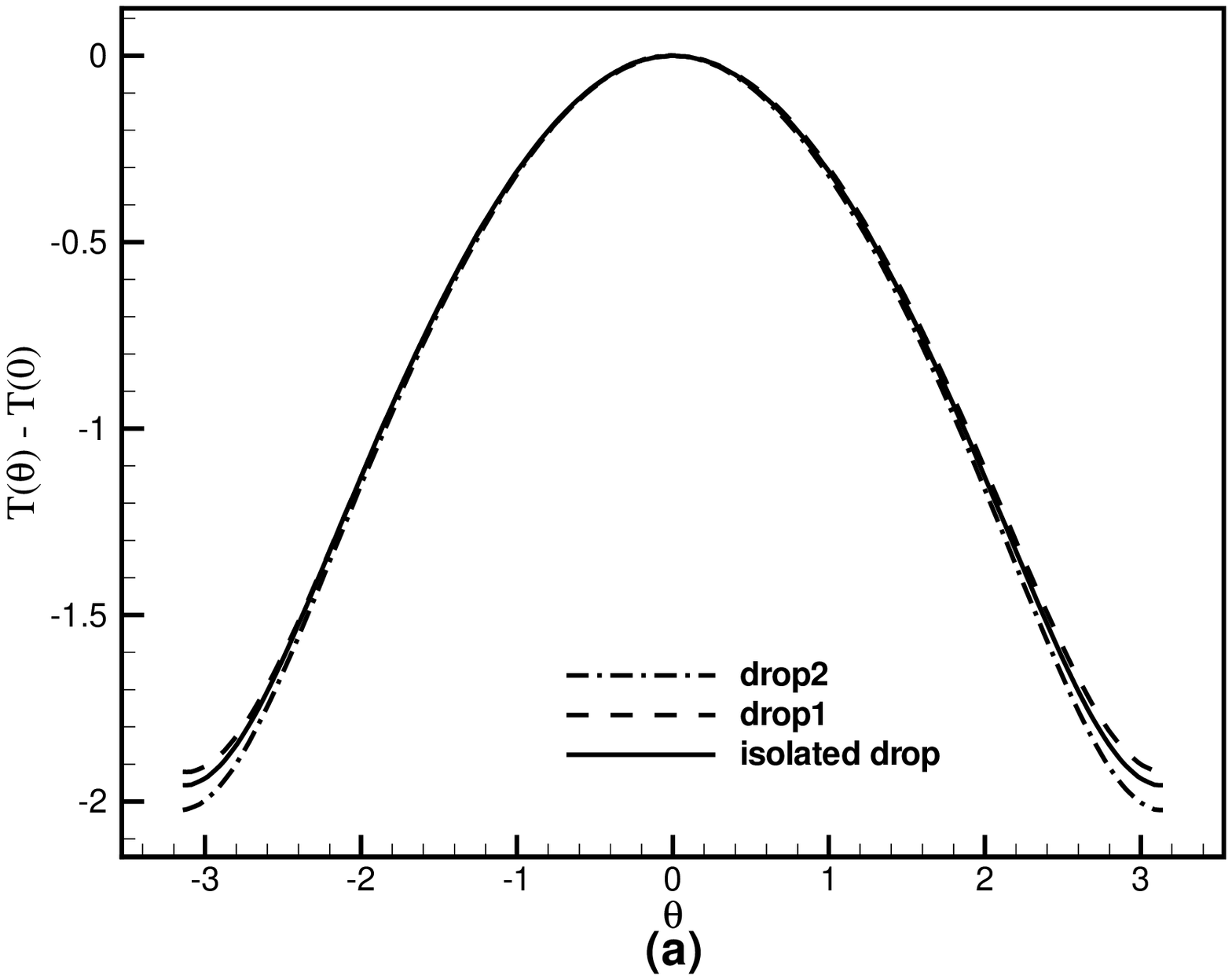}}\hspace{0.5in}
{\includegraphics[width=0.4\linewidth]{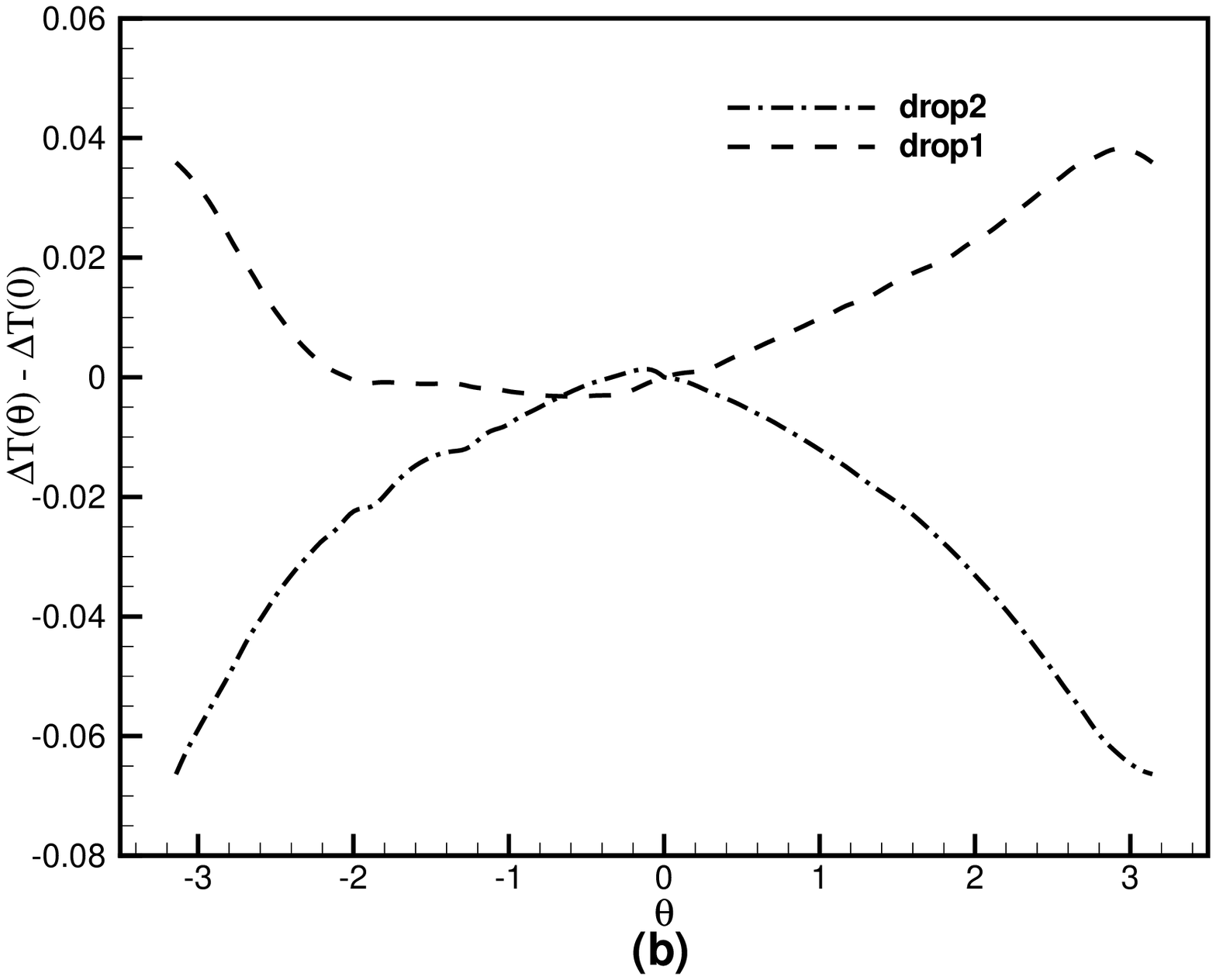}}
\caption{\label{fig:3d_Re1Ma20_t=60_surfaceT} The temperature distributing on the surface of drops in the $x=0$ plane at $t=60$.  $Re=1$, $Ma=20$, $S_{y0}=S_{z0}=1.35$ and $\Delta T(\theta)=T(\theta)-T_{iso}(\theta)$.}
\end{figure*}

The isotherms at $t=60$ are shown in Fig. \ref{fig:3d_Re1Ma20_t=60_iso_stream}. It can be seen that
the thermal convection of the lower left drop2 causes the bending of isotherms around the upper
right drop1. Compared with the isolated drop, the temperature gradient on the lower part of the
left drop2 is reduced, while that on the upper part of the right drop1 is enlarged. In order to get
a better understanding of the velocities in $z$ direction ($W$), we study the temperature
distributions at $t=60$ in the $x=0$ plane (Figs. \ref{fig:3d_Re1Ma20_t=60_surfaceT}). Compared
with the isolated drop, the temperature difference between front and rear stagnation points of
drop2 is larger, while that of drop1 is smaller.

The drop velocities in $y$ direction ($V$) are plotted in Fig. \ref{fig:3d_Re1Ma20_V} $(b)$. It is
found that drop1 is always trying to move away from drop2 in $y$ direction. Drop2 moves towards
drop1 in the beginning, but starts to move away since $t \approx 30$. Because $V_{drop2}$ is always
larger than $V_{drop1}$, the horizontal distance between two drops is increasing throughout the
simulation.

\subsection{Influence of the initial distance for the cases of $\Phi\neq0$}

\begin{figure*}
\begin{minipage}[t]{0.45\linewidth}
\begin{center}
{\includegraphics[width=0.9\linewidth]{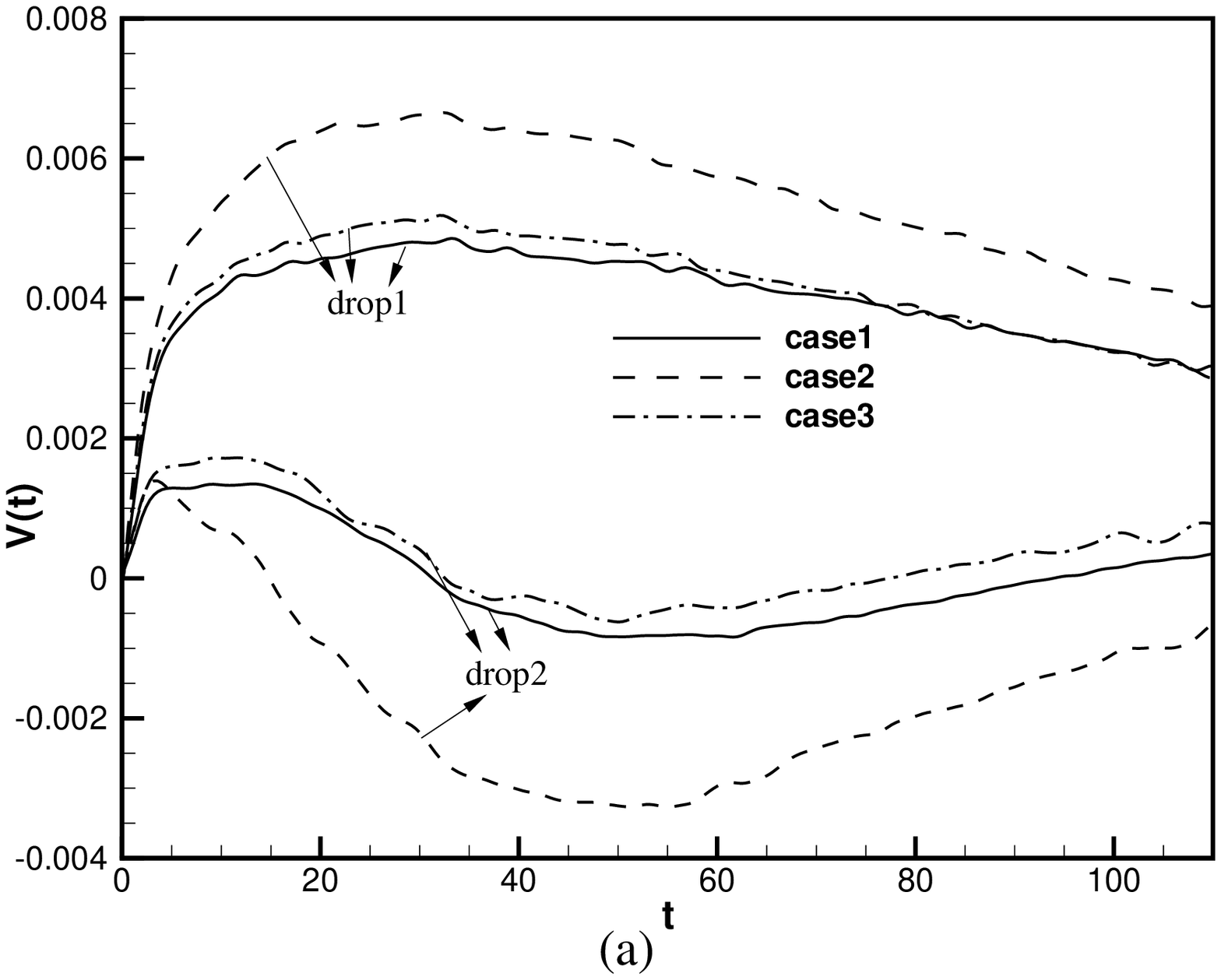}}
\end{center}
\end{minipage}
\hspace{0.2in}
\begin{minipage}[t]{0.45\linewidth}
\begin{center}
{\includegraphics[width=0.9\linewidth]{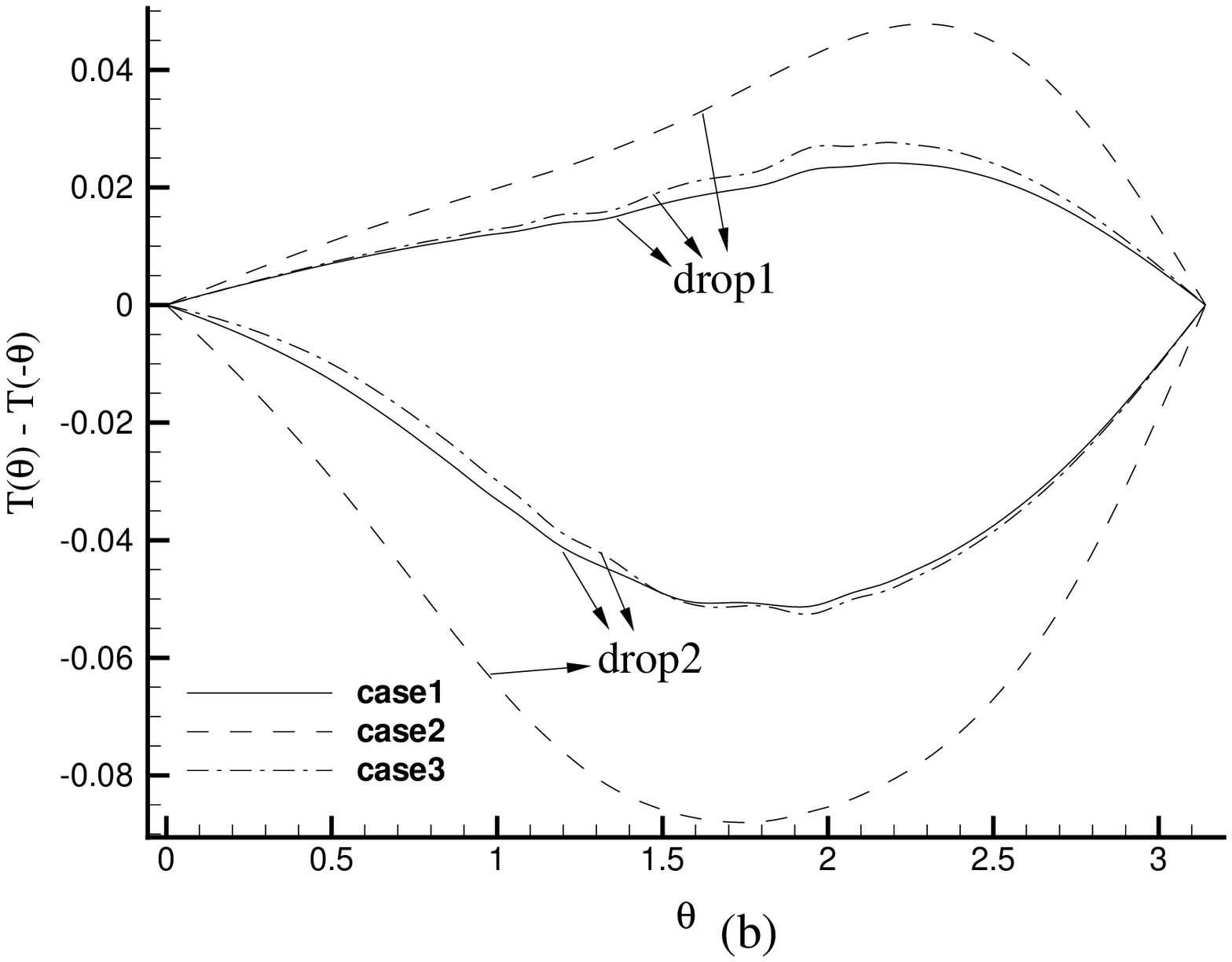}}
\end{center}
\end{minipage}
\caption{\label{fig:vardydz_Vy} (a) Time evolutions of the vertical velocities of the drops with various initial distances, (b) the temperature differences between the left and right sides of the drops.}
\end{figure*}

In this subsection, three sets of simulations with $Re=1$ and $Ma=20$ starting from different initial distances are studied:
\begin{enumerate}
   \item $(S_{y0}, S_{z0})=(1.35, 1.35)$;
   \item $(S_{y0}, S_{z0})=(1.1, 1.35)$;
   \item $(S_{y0}, S_{z0})=(1.35, 1.25)$.
\end{enumerate}
It is clear that the smaller the initial horizontal distance, the larger the velocities of two
drops in the $y$ direction, and the bigger the temperature differences between the left and right
sides of drops ($T(\theta)-T(-\theta)$, Fig. \ref{fig:vardydz_Vy}(b)).

With different Re and Ma numbers, the evolutions of distances between two drops for case1 are shown
in Fig. \ref{fig:3d_varReMa_dy_dz}. It can be seen that the two droplets separate very slowly when
$Re=1$ and $Ma=1$. When the Re number is increased, the two drops get close faster in the vertical
direction, while there are only trivial changes in separated distances for increasing Ma numbers.

Generally speaking, in the $z$ direction, the lower drop2 moves faster than the isolated drop,
while the upper drop1 moves slower than the isolated drop, and thus $S_z$ is decreasing throughout
any simulation in this subsection. If the simulation domain is big enough, drop2 would exceed drop1
in the $z$ direction and slow down to a velocity smaller than that of drop1. Then drop1 will start
to catch up with drop2, and so on. Eventually, both droplets will reach a steady migration state
when they are aligned horizontally, as indicated by Nas \emph{et al.}\cite{Nas:, Nas2:}.

\begin{table*}[h] 
\begin{center}
\begin{tabular}{p{3.1cm}|c|c|c|c|c}\toprule
Parameters & $Re=Ma=10^{-3}$ & $Ma=1, RE=1$ & $Ma=20, RE=1$ & $Ma=100, RE=1$ & rigid Spheres\\ \hline
$\Phi=0$ &$W_1+, W_2+$ &$W_1+, W_2+$ &$W_1+, W_2-$ &$W_1+, W_2-$ &$W_1+, W_2+$ \\ \hline

$\Phi=0.68, \pi/4, 0.82$ &--- &--- &$W_1-, W_2+$ &--- &$W_1+, W_2+$ \\ \hline

$\Phi=\pi/2$ &$W_1-, W_2-$ &--- &--- &--- &$W_1+, W_2+$ \\ \hline \hline
\end{tabular}
\caption{\label{tab:W1W2W} The velocities of drop1 ($W_1$) and drop2 ($W_2$) are compared with $W_{iso}$.
`$+$'/`$-$' means the velocity is bigger/smaller than $W_{iso}$. The velocities of rigid spheres  in Stokes flow are listed in the last row, the bigger/smaller velocity stands for the smaller/bigger resistance than that on the isolated rigid sphere.}
\end{center}
\end{table*}
\section{Conclusions}

In this paper, the interactions of two nondeformable droplets in thermocapillary motion are
studied. When the inertia and thermal convections are neglected, two vertically-placed drops will
move faster than the isolated drop, while two horizontally-placed drops will move slower. For the
finite $Ma$ number and $\Phi=0$, the leading drop moves faster than the isolated drop, while the
trailing drop migrates slower than the isolated drop due to the disturbed temperature field left by
the leading drop. When the two drops are closer, their interaction is stronger, but this intensive
interaction will not last long because the velocity difference of two drops is also big. Once there
is enough big gap between the two drops, they will migrate like the isolated drops. For the finite
$Ma$ number and $\Phi \neq 0$, the motions of two droplets are still limited in the $y-z$ plane.
The upper drop1 migrates slower while the lower drop2 migrates faster than the isolated drop, which
results in a smaller vertical distance and a bigger horizontal distance between the two drops. Tab.
\ref{tab:W1W2W} sums up the velocities of drop1 and drop2 in the $z$ direction ($W_1$, $W_2$)
studied in this paper.

Here, we only explore a few interacting mechanisms of two droplets with a limited number of parameters. A wider range of parameters as well as deeper physical explanations should be included in the future works.

\end{document}